\documentclass[multphys,vecphys]{svmult}

\usepackage{graphicx}
\usepackage{amssymb}
\usepackage{epsfig}
\def\Journal#1#2#3#4{{#1} {#2} (#3) #4}
\def\NPA{Nucl. Phys. A}
\def\NPB{Nucl. Phys. B}
\def\PLB{Phys. Lett.  B}
\def\PRL{Phys. Rev. Lett.}
\def\PRC{Phys. Rev. C}
\def\PRD{Phys. Rev. D}

\def\ZPC{Z. Phys. C}

\def\EPJC{Eur. Phys. J. C}

\def\ARNPS{Ann. Rev. Nucl. Part. Sci.}

\newcommand{\ud}{\mathrm{d}}
\newcommand{\be}{\begin{equation}}
\newcommand{\ee}{\end{equation}}

\newcommand{\AmS}{{\protect\the\textfont2
  A\kern-.1667em\lower.5ex\hbox{M}\kern-.125emS}}

\begin{document}


\title*{Ultrarelativistic nucleus-nucleus collisions and the quark-gluon plasma}

\author{A.~Andronic \and P.~Braun-Munzinger}
\institute{Gesellschaft f{\"u}r Schwerionenforschung, Darmstadt, Germany}
\maketitle


\vspace{-.4cm}
We present an overview of selected aspects of ultrarelativistic 
nucleus-nucleus collisions, 
a research program devoted to the study of strongly interacting matter
at high energy densities and in particular to the characterization of 
the quark-gluon plasma (QGP). 
The basic features of the phase diagram of nuclear matter, as currently
understood theoretically, are discussed.
The experimental program, carried out over a broad energy domain at various 
accelerators, is briefly reviewed, with an emphasis on  the global 
characterization of nucleus-nucleus collisions.  
Two particular aspects are treated in more detail:
i) the application of statistical models to a phenomenological description 
of particle production and the information it provides on the phase
diagram;
ii) the production of hadrons carrying charm quarks as messengers from
the QGP phase.



\vspace{.3cm}
{\it Go for the messes - that's where the action is.} 

S.Weinberg, Nature 426 (2003) 389

\section{Introduction}

Quantum Chromodynamics (QCD), the theory of strong interactions (see
\cite{aa:wil} for a recent review), predicts a phase transition from 
a state of hadronic constituents to a plasma of deconfined quarks and 
gluons, as the energy density exceeds a critical value.  
The opposite phase transition, from quarks and gluons to hadronic matter, 
took place about 10$^{-5}$ s after the Big Bang, the primeval event which 
is at the origin of our Universe.  
The core of the physics program of ultrarelativistic nucleus-nucleus 
collisions research \cite{aa:r1} is the study of the properties of strongly 
interacting matter at high energy density, in particular its phase diagram 
and the properties of quark-gluon plasma (QGP) \cite{aa:r2}.

Already in 1951 Pomeranchuk \cite{aa:pom} conjectured that a finite hadron
size implies a critical density, $n_c$, above which nuclear matter cannot 
be in a hadronic state.  In 1965, Hagedorn \cite{aa:hag} inferred that an
exponentially growing mass spectrum of hadronic states 
(observed up to masses of about 1.5 GeV) 
implies a critical temperature $T_c$ of the order of 200 MeV 
($\approx 2\cdot 10^{12}$~K). 
However, the elementary building blocks of QCD, the quarks and gluons
(carrying an extra quantum number called "color") have not been directly 
observed in experiments, although their fingerprints have been clearly 
identified in deep-inelastic collisions and jet production.
A fundamental property of QCD, the asymptotic freedom, unraveled by Gross, 
Wilczek, and Politzer in 1973 \cite{aa:gro}, implies that the attractive 
force (coupling) between quarks increases as a function of their 
separation.
Moreover, the confinement of quarks (and gluons) inside hadrons is 
another fundamental feature of QCD, although not fully understood yet.
Cabibbo and Parisi \cite{aa:cab} demonstrated already in 1975 that the 
exponential mass spectrum of hadronic states is a feature of any hadronic 
system which undergoes a second order phase transition with critical 
temperature $T_c$, since thermodynamical quantities exhibit singularities 
at $T_c$. 
This is realized in models that include "quark containment" \cite{aa:cab}, 
so is in agreement with QCD principles. 
Collins and Perry \cite{aa:col} demonstrated in the same year that 
asymptotically free QCD is also realized for large densities.
It is interesting to note that ref. \cite{aa:cab} contains the first sketch 
of a phase diagram of nuclear matter. 
The term quark-gluon plasma along with initial ideas about the space-time 
picture of hadronic collisions were first introduced by Shuryak \cite{aa:shu}.

\section{Theoretical background}

In the recent years, a successful effort to solve the QCD equations
numerically on a (space-time) lattice has brought deeper insight into 
the subject of phase transition(s) from hadronic to quark-gluon matter
\cite{aa:lqcd}.  It is not yet clear whether the transition is a true 
singular behaviour of thermodynamic variables or just a rapid crossover.

\begin{figure}[htb]
\vspace{.5cm}
\centering\includegraphics[width=.8\textwidth]{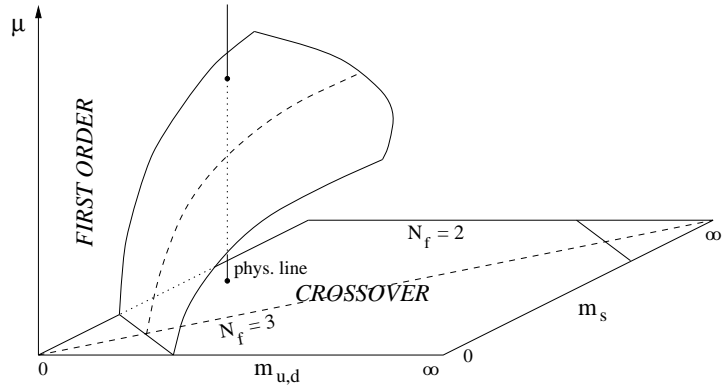}
\caption{Order of the phase transition in lattice QCD calculations 
in the variables quark masses (degenerate $u$, $d$ quarks and $s$ quark) 
and chemical potential (taken from ref. \cite{aa:lqcd}). 
} \label{aa:order}
\end{figure}

Fig.~\ref{aa:order} shows a map of the phase transition in the coordinates of
$u$,$d$ and $s$ quark masses and the chemical potential $\mu$ \cite{aa:lqcd}.
The surface, corresponding to a second order phase transition, is the border
between the regions of first order transition and crossover.  While the $u$
and $d$ quarks have small masses (of the order of a few MeV), the mass of the
$s$ quark is not well known, but is likely to be about 150 MeV. In this case
(so-called physical values of the quark masses, represented by the vertical
line in Fig.~\ref{aa:order}) the transition from QGP to a hadron gas is 
a crossover for small $\mu$ and reaches into the domain of the first order 
for large $\mu$, implying the existence of a critical point \cite{aa:crit}.
Up to now, experimental searches for such a critical point via enhanced
event-by-event fluctuations have not turned up any evidence \cite{aa:cerfl}.
Whether this means that all critical fluctuations are effectively damped 
by the phase transition or whether the transition is of first order, 
is currently an open question. 
It may also imply that the critical point is not reached in the energy range
studied up to now.

\begin{figure}[hbt]
\centering\includegraphics[width=.77\textwidth]{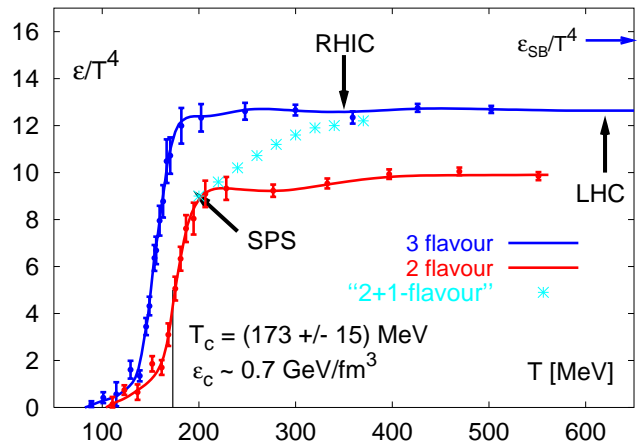}
\caption{Energy density as a function of temperature 
calculated with lattice QCD (taken from ref. \cite{aa:lqcd}).} 
\label{aa:eden}
\end{figure}

In any case, the transition to free quarks and gluons is illustrated by the
sudden increase of the energy density as a function of temperature, shown in
Fig.~\ref{aa:eden} for two and three degenerate flavors \cite{aa:lqcd}.  For
the 2-flavor case, the transition corresponds to a critical temperature
$T_c\simeq$170~MeV with critical energy density
$\varepsilon_c\simeq$0.7~GeV, while for the 3-flavor case $T_c$ is smaller by
about 20~MeV.  A result for the case of two degenerate flavors and a heavier
strange quark (physical values) is also included.

Other features of Fig.~\ref{aa:eden} can be understood be recalling 
fundamental results of thermodynamics of relativistic gases \cite{aa:cley}. 
The grand partition functions for fermions (particles and anti-particles) 
and bosons are:
\be
(T\ln Z)_f=\frac{g_fV}{12}\left(\frac{7\pi^2}{30}T^4+\mu^2T^2+
\frac{1}{2\pi^2}\mu^4\right) , \quad
(T\ln Z)_b = \frac{g_bV\pi^2}{90}T^4,
\label{eq:tlnz}
\ee
where $g_f$ and $g_b$ are the respective degeneracies (degrees of freedom).
The average energy, particle number and entropy densities and the pressure
are:
\be
\varepsilon=\frac{T}{V}\frac{\partial (T\ln Z)}{\partial T}+\mu n ,
\quad 
n=\frac{1}{V}\frac{\partial (T\ln Z)}{\partial \mu} ,
\quad P=\frac{\partial (T\ln Z)}{\partial V}, 
\quad
s=\frac{1}{V}\frac{\partial (T\ln Z)}{\partial T}
\ee
Using the thermodynamic relation: $\varepsilon=-P+Ts+\mu n$ one can easily 
establish the equation of state (EoS) of an ideal gas: $P=\varepsilon/3$.
Assuming that the hadronic world is composed of pions, $g_h$=3. 
For three colours and two spin values, for quarks and gluons one has 
$g_q$=12$N_f$ and $g_g$=16, respectively. $N_f$ is the number of flavours
(the lighter quarks $u$, $d$ and $s$ are the only ones relevant).
Consequently, at vanishing the chemical potential, the energy densities for 
the hadronic stage and for a gas of free quarks and gluons are, respectively:
\be
\varepsilon_h/T^4=\frac{\pi^2}{10}, \quad
\varepsilon_{qg}/T^4=(32+21N_f)\frac{\pi^2}{60}.
\ee
For 
$N_f$=3, $\varepsilon_{qg}/T^4$=15.6, denoted as the Stefan-Boltzmann limit, 
$\varepsilon_{SB}$, in Fig.~\ref{aa:eden}.
It is interesting to note that the calculated values are well below the 
values for non-interacting gases, indicating that the QGP is far from an 
ideal gas at temperatures as high as several times $T_c$.

An important (and not yet understood) result of lattice QCD calculations 
is that the critical temperatures for deconfinement and for chiral 
symmetry restoration ($T_\chi$) apparently coincide, although one 
might expect that $T_\chi\ge T_c$ \cite{aa:lqcd}. 

\begin{figure}[htb]
\centering\includegraphics[width=.63\textwidth]{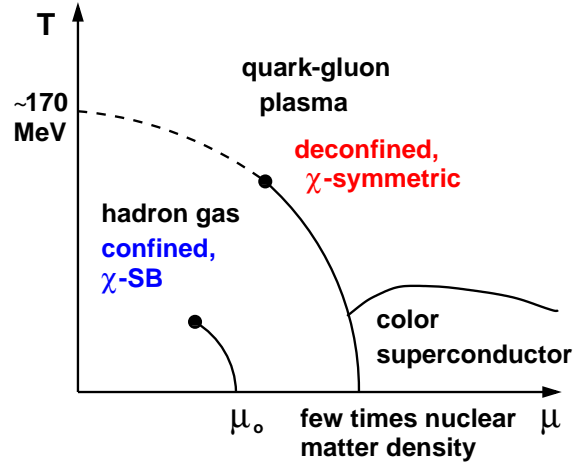}
\caption{Schematic phase diagram of nuclear matter (taken from ref. 
\cite{aa:lqcd}).} 
\label{aa:phase}
\end{figure}

A simple way to incorporate the two basic properties of QCD, asymptotic 
freedom and confinement, is achieved in the so-called (MIT) bag model 
\cite{aa:cley}.
It prohibits quarks ang gluons from existing  outside the bag (which can be 
any finite volume) by adding a shift from the physical vacuum into the QCD 
vacuum by an extra term in the partition function of the plasma phase:
$(T\ln Z)_{vac} = -BV$, where $B$ is the bag constant. 
It is easy to show that the EoS in this case becomes: $P=(\varepsilon-4B)/3$.
The phase transition trajectory in the $T-\mu$ plane can be constructed by
applying the Gibbs criteria for the phase transition: 
\be
P_h=P_{qg} , \quad \mu_h=\mu_{qg}(=3\mu_{q}), \quad T_h=T_{qg}=T_c.
\ee

A sketch of the present understanding \cite{aa:lqcd} of the phase diagram 
of strongly interacting matter is shown in Fig.~\ref{aa:phase} 
in the $T-\mu$ plane, for two light $u$ and $d$ quarks and a heavy
$s$ quark.
The lines mark the borders between the different phases of hadronic matter.
The dots mark the expected position of critical points, namely the
$T-\mu$ loci beyond which a first order phase transition is no longer
expected to take place.
Ground-state nuclear matter (atomic nuclei) corresponds to $\mu_0$=931~MeV
(bound nucleon mass) and $T$=0 and is well modeled as a liquid. 
The line starting at this point denotes the liquid-gas phase boundary
which is under study in low energy nucleus-nucleus collisions.
The region of high temperatures is the part which is being explored in 
ultrarelativistic nucleus-nucleus collisions. The exotic region of low 
temperatures and high densities (high $\mu$) is of relevance to astrophysical 
phenomena, but is rather likely to remain inaccessible to laboratory 
experiments. 

\section{Experimental program and global observables}

By colliding heavy ions at ultrarelativistic energies, one expects to create
matter under conditions that are sufficient for deconfinement \cite{aa:r1}.
A series of conferences (the so-called "Quark Matter" conferences, see ref. 
\cite{aa:qm} for the most recent of those) devoted to the subject has begun 
in 1980 (Bielefeld).

The experimental program has started at the CERN's Super Proton Synchrotron 
(SPS) and at Brookhaven's Alternating Gradient Synchrotron (AGS) in 1985.
The AGS program \cite{aa:ags}, carried out over a period of about 15 years 
by several experiments (E802/864,917 E810, E814/877, E864, E895)
is essentially completed.
The SPS program is just being concluded. Compelling evidence for the
production of a ``New State of Matter'',  has been produced in central Pb-Pb 
collisions \cite{aa:sps} studied by seven experiments: WA80/98, NA35/49, 
NA38/50/60, NA44 NA45/CERES, WA97/NA57, and NA52.
A vigorous research program, started with the first data taking in 2001, 
is on-going at the Relativistic Heavy Ion Collider (RHIC) at Brookhaven
National Laboratory (BNL) \cite{aa:rhic} with four experiments, BRAHMS, 
PHENIX, PHOBOS and STAR.
The Large Hadron Collider (LHC) will start operating at CERN in 2007 and
will provide (in addition to proton beams) heavy ion beams, which will
be used in the research program of the dedicated ALICE experiment as well
as by the ATLAS and CMS experiments \cite{aa:lhc}.
A dedicated fixed-target facility is planned at Gesellschaft f\"ur 
Schwerionenforschung (GSI), expected to be operational in 2012 \cite{aa:gsi}. 

The temporal evolution of a (central) nucleus-nucleus collision at 
ultrarelativistic energies is understood to proceed through the following 
stages: 
i) liberation of quarks and gluons due to the high energy deposited in the 
overlap region of the two nuclei; 
ii) equilibration of quarks and gluons; 
iii) crossing of the phase boundary and hadronization; 
iv) freeze-out.  
Interesting experimental information is contained in the study of the 
distributions of (mostly charged) hadrons after freeze-out.
Whether any information on the phase transition can be gleaned from these 
investigations will be discussed below. Clearly, given the short timescales 
of a nucleus-nucleus collision and the small volume involved 
(lattice QCD calculations discussed in the previous section are for bulk) 
the reconstruction of the various stages of the collision is a difficult task.
It is consequently of particular relevance to find experimental observables 
which carry information (preferentially) from one particular stage, in 
particular about the QGP phase. Specific probes of QGP have been proposed
\cite{aa:bj,aa:qgp} and are currently being studied experimentally: 
i) direct photons \cite{aa:photo};
ii) low-mass dileptons \cite{aa:dilep};
iii) strangeness \cite{aa:strange}; 
v) charmonium suppression \cite{aa:r2}; 
vi) jet-quenching \cite{aa:jets};
vii) fluctuations \cite{aa:cerfl,aa:fluct}.

As shown in the next section, the study of hadron multiplicities in a
statistical model is a unique way to provide experimental information on 
the QCD phase diagram \cite{aa:therm}. Other global observables, like the
distribution of particles over momentum space, collective flow, and the
measurements of effective source sizes via particle interferometry, have 
also been studied in detail. In particular their energy evolution is of 
relevance and is briefly examined in the following ($\sqrt{s_{NN}}$ is 
the total center-of-mass energy per nucleon pair).

\begin{figure}[htb]
\vspace{-1cm}
\begin{tabular}{cc}
\begin{minipage}{.58\textwidth}
\centering\includegraphics[width=1.15\textwidth]{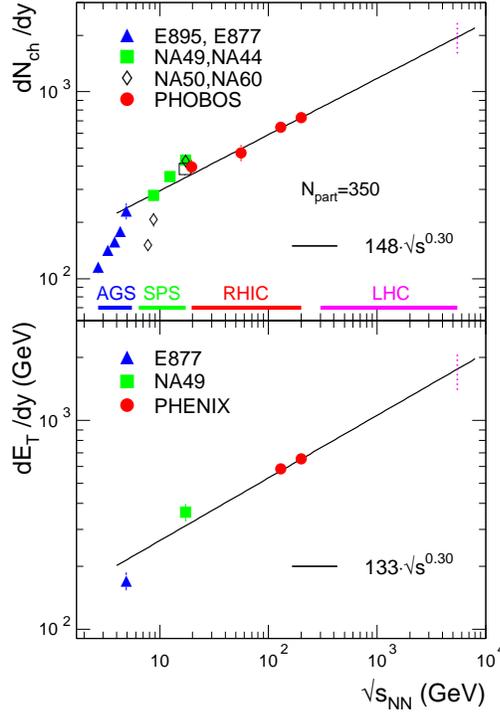}
\end{minipage} & \begin{minipage}{.38\textwidth}
\caption{Excitation function of global observables in central 
central nucleus-nucleus collisions ($N_{part}$=350).  
The experimental values for particle rapidity density, $\ud N_{ch}/\ud y$ 
\cite{aa:e895,aa:e877,aa:na49,aa:na44,aa:na50ch,aa:na60,aa:phobos}
(upper panel) and transverse energy rapidity density, $\ud E_T/\ud y$ 
\cite{aa:e877,aa:et} (lower panel) at midrapidity are plotted as symbols.
The lines are a power law dependence arbitrarily scaled.
The thick horizontal lines mark the energy range of the various accelerators.
The dotted line marks the full LHC energy for Pb--Pb collisions 
($\sqrt{s_{NN}}$=5.5 TeV).} 
\label{aa:nch}
\end{minipage} \end{tabular}
\end{figure}

In Fig.~\ref{aa:nch} we present a compilation of experimental data on charged
particle rapidity density distributions, $\ud N_{ch}/\ud y$, and transverse
energy rapidity density, $\ud E_T/\ud y$, at midrapidity\footnote{Midrapidity
is the rapidity of center-of-mass system; rapidity is defined as 
$y=0.5\ln[(E+p_z)/(E-p_z)]$, where $p_z$ is the longitudinal component of 
the particle momentum and $E$ is the energy.}. 
The values are for central collisions (average value of the number of 
participant nucleons in the fireball, $N_{part}$=350, which roughly corresponds
to the 5\% most central collisions) in the energy range from AGS up to 
RHIC.\footnote{A constant Jacobian of 1.1 has been used to convert the 
$\ud X/\ud\eta$ data to $\ud X/\ud y$. \\ $\eta=-\ln[\mathrm{tan}(\theta/2)]$ 
is the pseudo-rapidity ($\theta$ is the polar angle of a given particle).}
The continuous lines are $(\sqrt{s_{NN}})^{0.3}$ dependences, arbitrarily
normalized. These power-law dependences describe the measurements quite well
starting from the top AGS energy ($\sqrt{s_{NN}}\simeq$5~GeV).  This may
suggest that some aspects of the underlying physics are similar over all this
energy domain.  Note that the SPS data (NA49) seem to slightly deviate from 
the power-law behaviour. Also, at the lower SPS energies there is an apparent
disagreement between NA49 and NA50/NA60 data.
The $(\sqrt{s_{NN}})^{0.3}$ dependences allow for simple, experimentally-based,
extrapolations to the LHC energy (of course, surprises are eagerly awaited).
We note that power-law dependences are predicted by the (QCD-inspired)
saturation model \cite{aa:esk}, but they are steeper (exponent 0.41, 
in case of $\ud N_{ch}/\ud y$).
The steep decrease of particle multiplicities towards the lower end of 
the AGS energy range reflects mainly the threshold in the overall particle 
production, but may indicate a change in physics as well.
The average transverse energy per charged particle has a nearly constant
value of 0.8-0.9 GeV all the way from top AGS to RHIC energies.

The initial energy density, $\varepsilon$, and the net baryon density, 
$n_{baryon}$ produced in a (central) heavy ion collision can be calculated 
from the measured transverse energy ($\ud E_T/\ud\eta$) and net baryon 
(${\ud N_{b-\bar b}}/{\ud\eta}$) densities, respectively, in the so-called 
"Bjorken-scenario" \cite{aa:bj}.
This assumes self-similar (Hubble-like) homogeneous (hydrodynamical) expansion
of the fireball in the longitudinal (beam) direction.
The resulting relations are:
\be
\varepsilon = \frac{1}{A_{T}}\frac{\ud E_T}{\ud\eta}\frac{\ud\eta}{\ud z},
\quad
n_{baryon} = \frac{1}{A_{T}}\frac{\ud N_{b-\bar b}}{\ud\eta}
\frac{\ud\eta}{\ud z}
\label{eq:bj}
\ee
where $A_{T}$ is the transverse area of the fireball ($A_{T}=$154~fm$^2$
for a head-on Au-Au collision).
In the above equations the only unknown parameter is the formation time 
(the time for establishing the equilibrium), $\tau$ ($\ud\eta/\ud z=1/\tau$), 
which is usually taken to be 1~fm/c, although it is expected to decrease 
as a function of the energy. In this sense, the values obtained using 
Eq.~\ref{eq:bj} are conservative estimates for most of the energy range 
spanned by the experiments.

\begin{table}[htb]
\caption{Measured and deduced quantities at AGS, SPS, and RHIC for 
central nucleus-nucleus collisions. For the LHC case the values are 
extrapolations (see text).}
\label{aa:tab0}
\begin{tabular}{c|cccc} 
Machine                 &  AGS    & SPS  &  RHIC &  LHC  \\ \hline
$\sqrt{s_{NN}}$ (GeV)   &  4.9    & 17.3 &  200  &  5500  \\ \hline
$\ud E_T/\ud \eta$ (GeV)    & 192 &  363  & 625  &  1800 ?  \\ 
$\ud N_{b-\bar b}/\ud \eta$ & 170 &  100  &  25  &  $\sim$0 ?  \\  \hline
$\varepsilon$ (GeV/fm$^3$)  & 1.2 &  2.4  & 4.1  &  11.6 ? \\
$n_{baryon}$ (fm$^{-3}$)     & 1.1 &  0.65 & 0.17 &  ? \\
\end{tabular}
\end{table}

The maximum nucleon-nucleon center-of-mass energy ($\sqrt{s_{NN}}$) and 
the corresponding measured and calculated (using Eq.~\ref{eq:bj}) energy 
and baryon densities are listed in Table~\ref{aa:tab0} for the various 
accelerator regimes.
The results for LHC are extrapolations based on the $(\sqrt{s_{NN}})^{0.3}$ 
dependence discussed above.

The achieved densities are obviously very much larger than  those 
inside a normal Pb nucleus: $\varepsilon_0=0.15$~GeV/fm$^3$ and 
$n_0=0.16$/fm$^3$.
The estimates of the energy densities are for all the energy range above
the critical energy density at $\mu_b=0$ 
($\varepsilon_{c}\simeq0.7$~GeV/fm$^3$),
indicating that the conditions for the QGP formation  likely  
have been achieved in the experiments.

\begin{figure}[htb]
\vspace{-.8cm}
\centering\includegraphics[width=.74\textwidth,height=.66\textwidth]{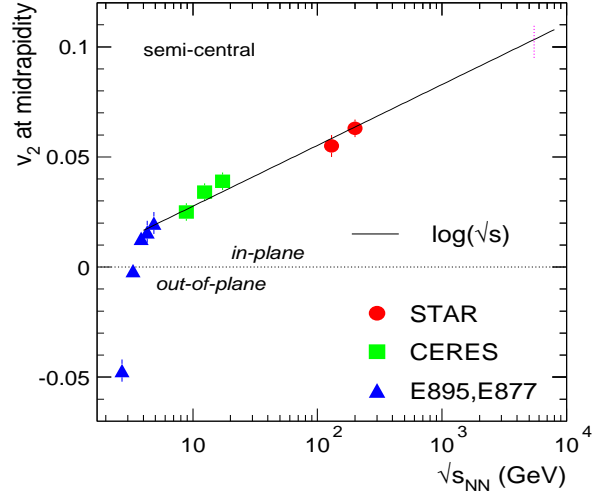}
\caption{Excitation function of elliptic flow. Protons are considered up
to SPS energies and all charged particles at RHIC. 
The line is a $\log(\sqrt{s_{NN}})$ dependence arbitrarily normalized.} 
\label{aa:v2} 
\end{figure}

In Fig.~\ref{aa:v2} we present the excitation function of elliptic flow
\cite{aa:elli} for semi-central collisions (for which this observable has 
a maximum as a function of centrality). 
This observable is characterized by the second order Fourier
coefficient $v_2=\langle\cos(\phi)\rangle$, where $\phi$ is the azimuthal 
angle with respect to the reaction plane (defined by the impact parameter 
vector and the beam direction).
It reflects the initial geometry of the overlap region and its 
pressure gradients. A transition from out-of-plane (also called "squeeze-out",
$v_2<$0) to in-plane ($v_2>$0) preferential particle emission is seen in
the energy domain of the  AGS. This is the result of a combined effect of the
violence of the expansion and of the shadowing of the spectator matter, 
which, at these energies is still present in the vicinity of the fireball.
A striking correlation of this transition with the sharp increase of particle
multiplicity at midrapidity (seen in Fig.~\ref{aa:nch}) is evident.
From top AGS energy up to RHIC the $v_2$ values increase steadily and 
are also well described by a $\log(\sqrt{s_{NN}})$ dependence.
An early observation at RHIC was that the $v_2$ values are reaching the 
hydrodynamical \cite{aa:hydro} limits which is an indication of an early 
equilibration of the fireball. 
It is likely that QGP is the only way to achieve such a fast equilibration.
It is thus an interesting question whether elliptic flow at LHC will follow
the $\log(\sqrt{s_{NN}})$ trend or will flatten at the RHIC values.

\begin{figure}[htb]
\vspace{-1cm}
\centering\includegraphics[width=.74\textwidth,height=.66\textwidth]{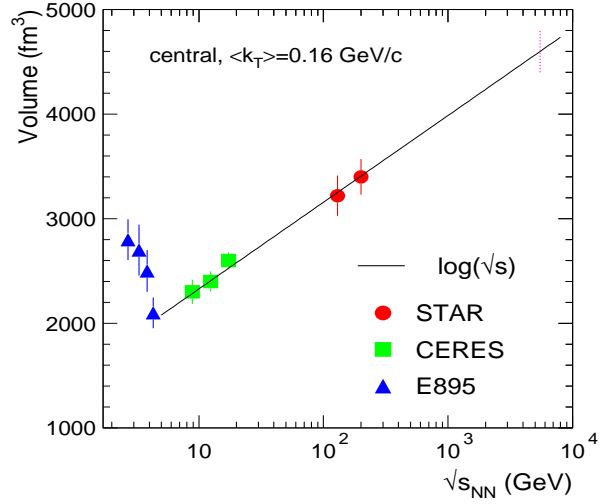}
\caption{Excitation function of the freeze-out volume extracted from  
pion HBT correlations. The line is a $\log(\sqrt{s_{NN}})$ dependence
arbitrarily normalized.} 
\label{aa:vol} 
\end{figure}

\vspace{-1.2cm}
In Fig.~\ref{aa:vol} we show the energy dependence of the volume of the
fireball as extracted from pion Hanbury~Brown-Twiss (HBT) correlations
\cite{aa:hbtrev}. Again, a strikingly different behavior is seen at the 
lowest energies compared to top AGS and above, for which a 
$\log(\sqrt{s_{NN}})$ dependence describe the measurements \cite{aa:hbt} 
well.  This non-monotonic behavior can be understood \cite{aa:hbtcer} 
quantitatively as the result of an universal pion freeze-out at a critical 
mean free path $\lambda_f\simeq$1~fm, independent of energy.  
It is worth mentioning that the smooth evolution of the source size in 
the energy range of top AGS to RHIC is an indication that no first order 
phase transition, associated with supercooling and explosive
expansion, is visible in hadronic observables in this energy domain. 
Also, the measured source sizes at RHIC are well below hydrodynamic 
predictions.

\section{Particle yields and their statistical description}

The equilibrium behavior of thermodynamical observables can be
evaluated as an average over statistical ensembles. The equilibrium
distribution is thus obtained by an average over all accessible
phase space. Furthermore, the ensemble corresponding to
thermodynamic equilibrium is that for which the phase space
density is uniform over the accessible phase space. In this sense,
filling the accessible phase space uniformly is both a necessary
and a sufficient condition for equilibrium.
We restrict ourselves here on the basic features and essential
results of the statistical model approach. A complete survey of 
the assumptions and results, as well as of the relevant references, 
is available in ref.~\cite{aa:therm}.

The basic quantity required to compute the thermal composition of
particle yields  measured in heavy ion collisions is the
partition function $Z(T,V)$. In the grand canonical (GC) ensemble,
for particle $i$ of strangeness $S_i$, baryon number $B_i$, electric 
charge $Q_i$ and spin-isospin degeneracy factor $g_i=(2J_i+1)(2I_i+1)$,
the partition function is:
\be 
\ln Z_i ={{Vg_i}\over {2\pi^2}}\int_0^\infty \pm p^2\ud p \ln [1\pm 
\exp (-(E_i-\mu_i)/T)] 
\label{eq:part}
\ee
with (+) for fermions (like baryons, made of 3 quarks) and (--) for bosons 
(like mesons, made of quark-antiquark pairs). Note that the partition 
functions introduced in Eq.~\ref{eq:tlnz} are for massless particles, 
for which the analytic integration of Eq.~\ref{eq:part} can be performed.
The particle density is:
\be
n_i=N/V=-\frac{T}{V}\frac{\partial\ln Z_i}{\partial\mu}=\frac{g_i}{2 \pi^2} 
\int_0^\infty \frac{p^2 \ud p}{\exp[(E_i-\mu_i)/T] \pm 1} 
\ee
$T$ is the temperature and $E_i =\sqrt {p^2+m_i^2}$ is the total energy.
$\mu_i = \mu_b B_i+\mu_S S_i+\mu_{I_3} I_{3i}$ is the chemical potential, 
with $\mu_B$, $\mu_S$, and $\mu_Q$ the chemical potentials related to baryon 
number, strangeness and electric charge, respectively, which ensure the 
conservation (on average) the respective quantum numbers: 
i) baryon number: $V\sum_i n_i B_i = Z+N$; 
ii) strangeness: $V \sum_i n_i S_i = 0$;
iii) charge:  $V \sum_i n_i I_{3i} = \frac{Z -N}{2}$.
This leaves $T$ and the baryochemical potential $\mu_b$ as the only parameters 
of the model. In practice, however, the volume determination may be subject
to uncertainties due to incomplete stopping of the colliding nuclei. Due to
this reason, the most convenient way to compare with measurements is to use 
particle ratios.

The interaction of hadrons and resonances is usually included by 
implementing a hard core repulsion of Van der Waals--type via
an excluded volume correction. This is implemented in an iterative
procedure according to:
\be P^{excl.} (T,\mu)= P^{id.gas}(T,\hat{\mu}); \qquad 
\hat{\mu} = \mu - V_{eigen} P^{excl.}(T,\mu) \ee
where $V_{eigen}$ is calculated for a radius of 0.3~fm, considered identical
for all particles.

The grand canonical ensemble is of course the simplest realization
of a statistical approach and is suited for large systems, with
large number of produced particles.
However, for small systems (or peripheral nucleus-nucleus  collisions) and
for low energies in case of strangeness production, a canonical ensemble (C)
treatment is mandatory. It leads to severe phase space reduction for 
particle production (so-called ``canonical suppression'').
Within this approach, particle production in e$^+$e$^-$ collisions 
has been successfully described, albeit with an additional heuristic
strangeness suppression factor. 
It has been shown 
that the density of particle $i$ with strangeness $S$ calculated in the 
canonical approach, $n_{i}^{C}$, is related to the grand canonical value, 
$n_{i}^{GC}$, as: $n_{i}^{C} = n_{i}^{GC} F_S$, with $F_S={I_{S}(x)}/{I_0(x)}$.
The argument of the Bessel function of order $S$ is the total yield of 
strange and antistrange particles.
For central Pb-Pb (Au-Au) collisions, the canonical suppression is negligible 
for all strange particle species already for the highest AGS energy 
($\sqrt{s_{NN}}\simeq$5~GeV) but is sizeable for the lowest energy considered
in the following, $\sqrt{s_{NN}}=$2.7~GeV (corresponding to the beam energy 
of 2~GeV/n), for which $F_1\simeq$2, $F_2\simeq$8. 

\begin{figure}[htb]
\vspace{-1cm}
\centering\includegraphics[width=.83\textwidth]{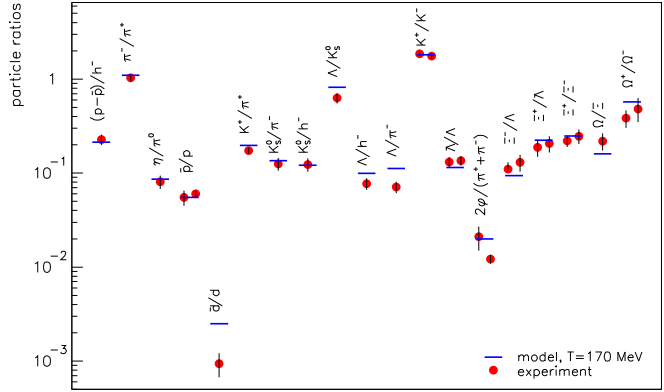}
\caption{Fit of particle ratios for Pb-Pb collisions at SPS (158 GeV/c).
The measurements are the symbols, the thermal fit values are the lines.} 
\label{aa:sps}
\end{figure}

In Fig.~\ref{aa:sps} we present the result of a thermal fit of the measured 
particle ratios for Pb-Pb collisions at 158 GeV/nucleon beam energy.
The values $T$=170$\pm$5~MeV and $\mu_b$=255$\pm$10~MeV
are the free parameters. 
The reduced $\chi^2$ (excluding $\phi$ and d) is 2.0, of which the largest 
contribution comes from the ratios $\Lambda/\pi$, $\Lambda$/h$^-$ and
$\Lambda$/K$^0_s$, possibly due to weak decays feeding.

\begin{figure}[htb]
\centering\includegraphics[width=1.03\textwidth]{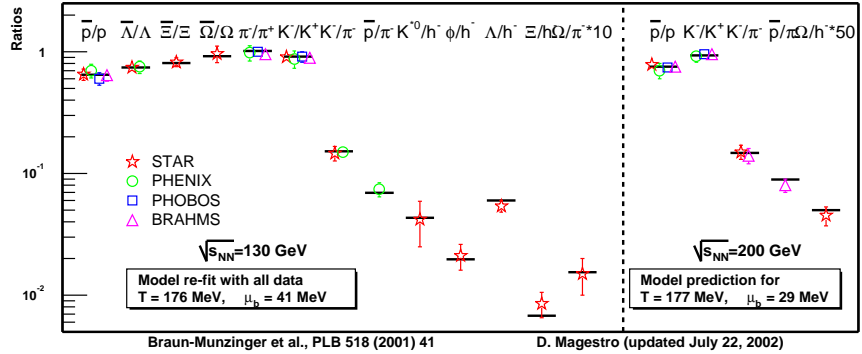}
\caption{Fit of particle ratios for Au-Au collisions at RHIC.
The measurements are the symbols, the thermal model values are the lines.} 
\label{aa:rhicth}
\end{figure}

The thermal fits of particle ratios for the RHIC energies ($\sqrt{s_{NN}}$=130 
and 200 GeV) are shown in Fig.~\ref{aa:rhicth}.
The obtained values for (T,$\mu_b$) are (174$\pm7$,46$\pm5$) MeV and 
(177$\pm7$,29$\pm 6$) MeV, respectively, with reduced $\chi^2$ values of 
0.8 and 1.1.

We mention here that the measured enhancement of strange hyperons ($\Lambda$,
$\Xi$, $\Omega$) at SPS in central Pb-Pb collisions with respect to pBe and 
pPb (a factor of 20 enhancement in case of $\Omega$) can be understood
quantitatively not as an enhancement in central Pb-Pb but as a  
suppression in pBe/pPb with respect to central Pb-Pb.
It is also important to note that, at RHIC, the transverse momentum spectra 
can be well described in a thermal approach, with two additional (size) 
parameters \cite{aa:bro}.
At AGS, the measured yields of light nuclei ($A\le$7) are well explained 
by the thermal model \cite{aa:pbm2}.

\begin{figure}[htb]
\vspace{-1.cm}
\begin{tabular}{cc}
\begin{minipage}{.59\textwidth}
\centering\includegraphics[width=1.1\textwidth, height=1.3\textwidth]{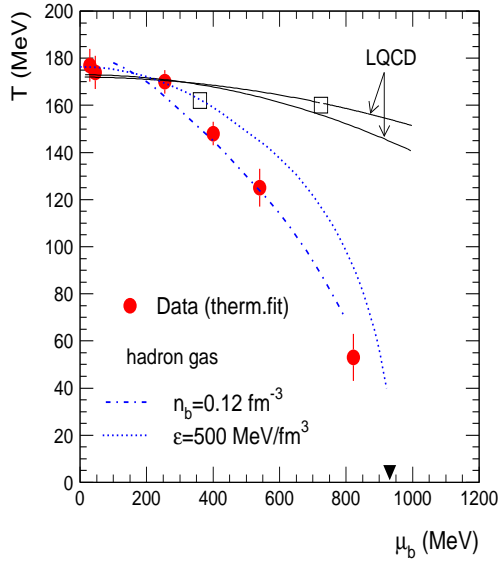}
\end{minipage} & \begin{minipage}{.37\textwidth}
\caption{The phase diagram of nuclear matter in the $T-\mu_b$ plane. 
The dots represent the extracted values from thermal fits to measured 
particle ratios.
The trajectories of freeze-out for a hadron gas at constant energy density 
($\varepsilon$=500~MeV/fm$^3$) and at constant baryon density 
($n_b$=0.12~fm$^{-3}$) are shown by the dotted and dash-dotted lines, 
respectively.
The phase boundary from lattice QCD (LQCD) calculations is shown with 
continuous lines. 
The open squares indicate the critical point with two different
inputs for calculations \cite{aa:crit}.
} 
\label{aa:phase2}
\end{minipage} \end{tabular}
\end{figure}

$T$ and $\mu_b$ were determined for other energies (SPS at 40 GeV/n, AGS 
at 10.8 GeV/n and for 1 GeV/n Au-Au collisions at SIS) with a similar fitting 
procedure, although using in most cases fewer available measured ratios
\cite{aa:therm}.
The resulting values are shown in a phase diagram of hadronic  matter
\cite{aa:pbm2} in Fig.~\ref{aa:phase2}, together with calculations of 
freeze-out trajectories or a hadron gas at constant energy density and 
at constant baryon density.
This latter case, corresponding to $n_b$=0.12~fm$^{-3}$, does reproduce
well the freeze-out points extracted from the data.
Another observation is that the freeze-out points lie on a curve corresponding
to an average energy $\langle E\rangle$ per average number of hadrons 
$\langle N\rangle$ of approximately 1~GeV.
We have mentioned above that an universal pion freeze-out corresponding 
to a mean free path of about 1~fm has been derived from HBT source size 
measurements \cite{aa:hbtcer}.

An important observation about the phase diagram is that, for the top SPS 
energy and above, the thermal parameters are (implying hadron yields frozen)
at the phase boundary, as known from lattice QCD calculations \cite{aa:crit}.
A natural question though is how is equilibrium achieved? 
Considerations about collisional rates and timescales of the hadronic 
fireball expansion \cite{aa:sto} imply that at SPS and RHIC the equilibrium
cannot be established in the hadronic medium and that it is the phase 
transition which drives the particles densities and ensures chemical 
equilibrium.

In a recent paper \cite{aa:pbmx} many body collisions near $T_c$ were
investigated as a possible mechanism for the equilibration.  There it is
argued that because of the rapid density change near a phase transition such
multi-particle collisions provide a natural explanation for the observation 
of chemical equilibration at RHIC energies and lead to $T=T_c$ to within
an accuracy of a few MeV. Any scenario with $T$ substantially smaller
than $T_c$ would require that either multi-particle interactions dominate 
even much below $T_c$ or that the two-particle cross sections are larger 
than in the vacuum by a high factor. 
Both of the latter hypothesis seem unlikely in view of the rapid density 
decrease. The critical temperature determined from RHIC for $T\approx T_c$ 
coincides well with lattice estimates \cite{aa:lqcd} for $\mu=0$, as 
discussed above.  The same arguments as discussed here for RHIC energy 
also hold for SPS energies: it is likely that also there the phase 
transition drives the particle densities and ensures chemical equilibration.

We note that thermal models have also been used \cite{aa:beca} to describe
hadron production in e$^+$e$^-$ and hadron-hadron collisions, leading to
temperature parameters close to 170 MeV. This suggests that hadronization
itself can be seen as a prethermalization process. However, to account for the
strangeness undersaturation in such collisions, multi-strange baryons can only
be reproduced by introducing a strangeness suppression factor of about 0.5,
leading to a factor of 8 suppression of $\Omega$ baryons.  This
non-equilibrium feature, also visible in the momentum distributions of the
produced particles, is most likely due to the ''absence'' of multi-particle
scattering since the system is not in a high density phase due to a phase
transition.

\begin{figure}[htb]
\vspace{-1.cm}
\hspace{-.6cm}
\begin{tabular}{cc}
\begin{minipage}{.49\textwidth}
\centering\includegraphics[width=1.2\textwidth,height=1.3\textwidth]{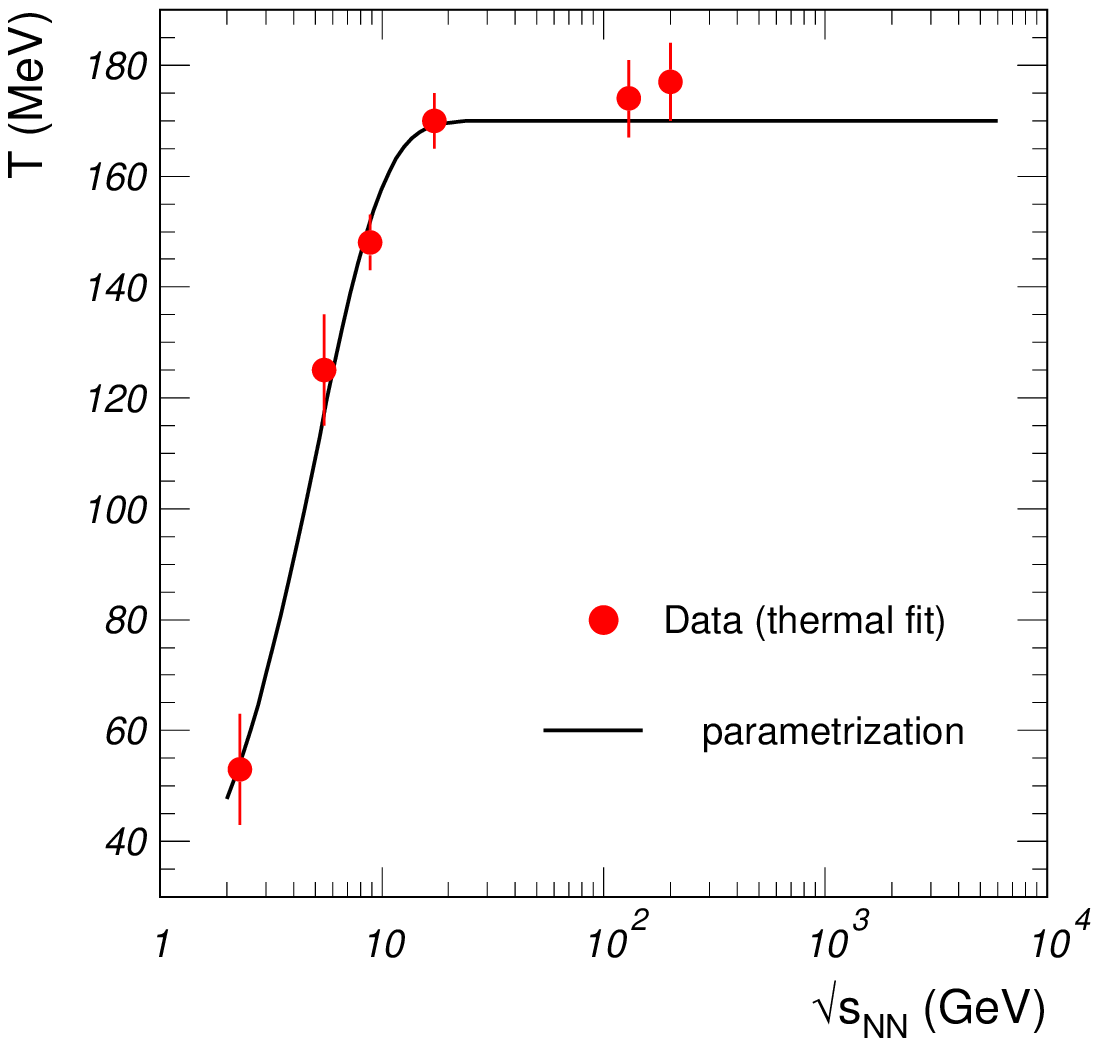}
\end{minipage} & \begin{minipage}{.49\textwidth}
\centering\includegraphics[width=1.2\textwidth,height=1.3\textwidth]{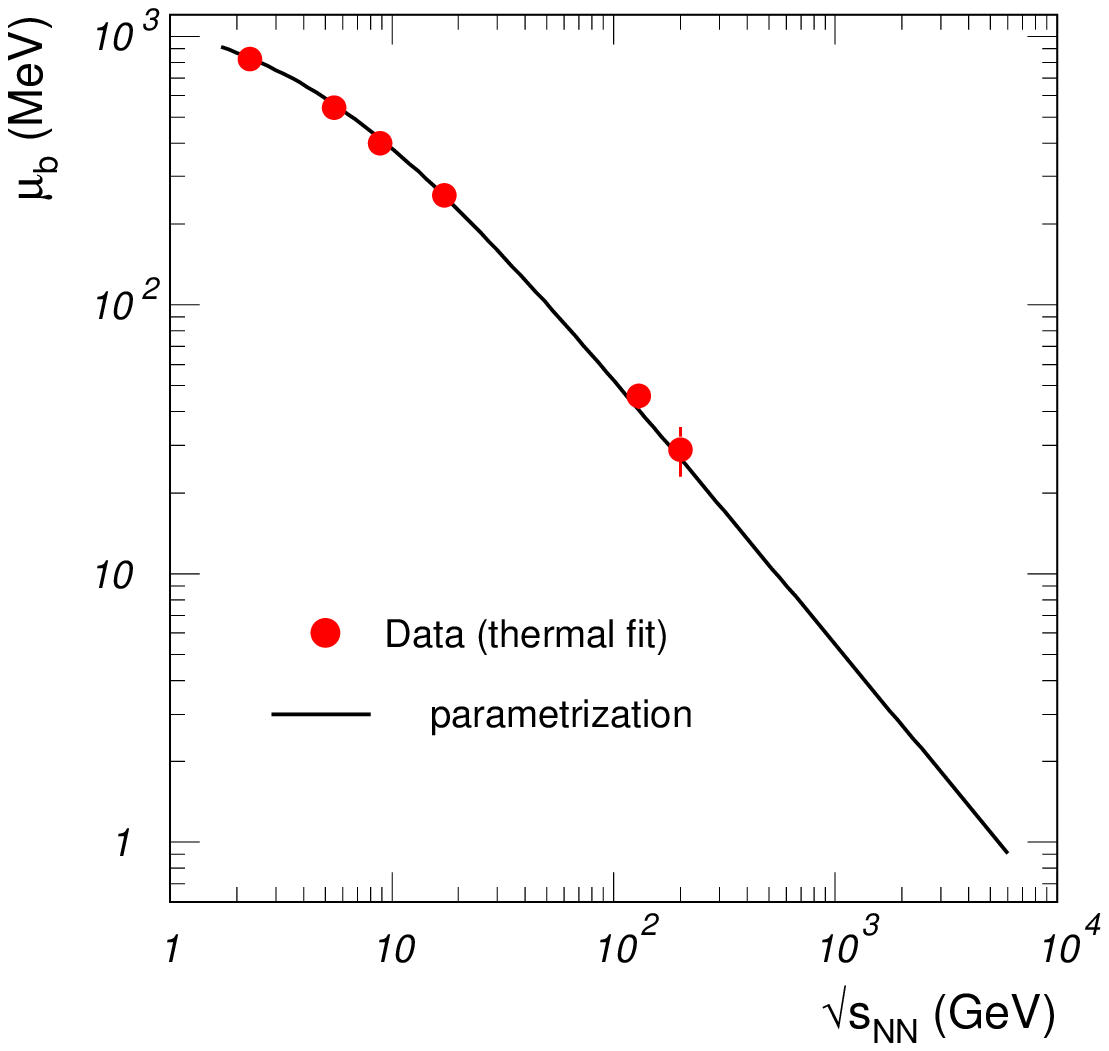}
\end{minipage} \end{tabular}
\caption{Energy dependence of the thermal parameters $T$ and $\mu_b$.
The symbols are the values extracted from experimental data, the lines are 
parametrizations (see text).
} 
\label{aa:tmus}
\end{figure}

The energy dependence of the extracted $T$ and $\mu_b$ values is presented 
in Fig.~\ref{aa:tmus}. The lines are parametrizations
that allow for extrapolating the parameters up to the LHC energy.
For $\mu_b$ the following parametrization has been used \cite{aa:therm}:
\be
\mu_b=1270[{\rm MeV}]/(1+\sqrt{s_{NN}}[{\rm GeV}]/4.3), \label{eq:mu_b}
\ee
while $T$ has been described with a Fermi-like function.
For both cases the parametrizations describe well the extracted values
over all the energy range.

\begin{figure}[htb]
\vspace{-.6cm}
\hspace{-.6cm}
\begin{tabular}{cc}
\begin{minipage}{.49\textwidth}
\centering\includegraphics[width=1.2\textwidth]{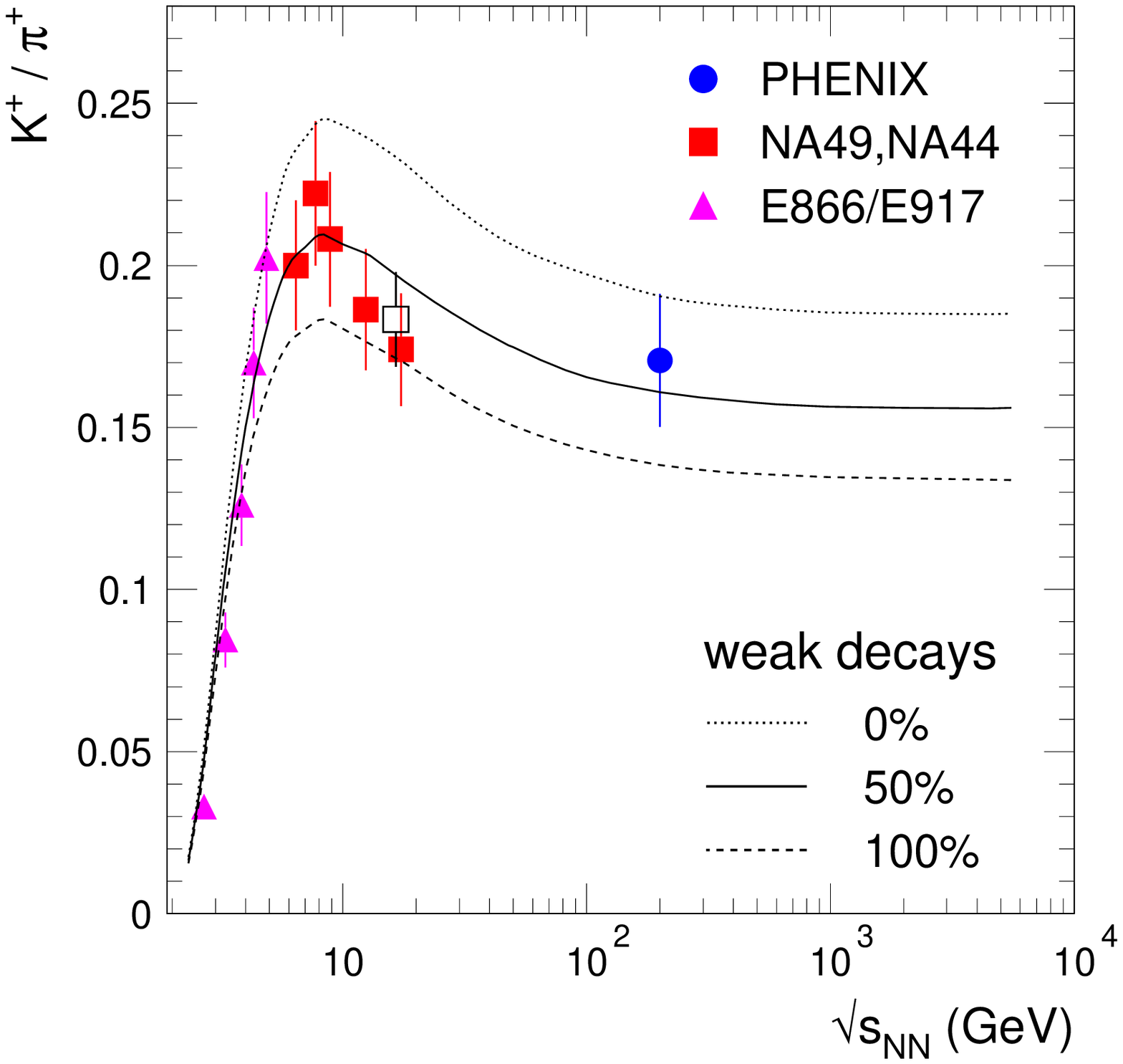}
\end{minipage} & \begin{minipage}{.49\textwidth}
\centering\includegraphics[width=1.2\textwidth]{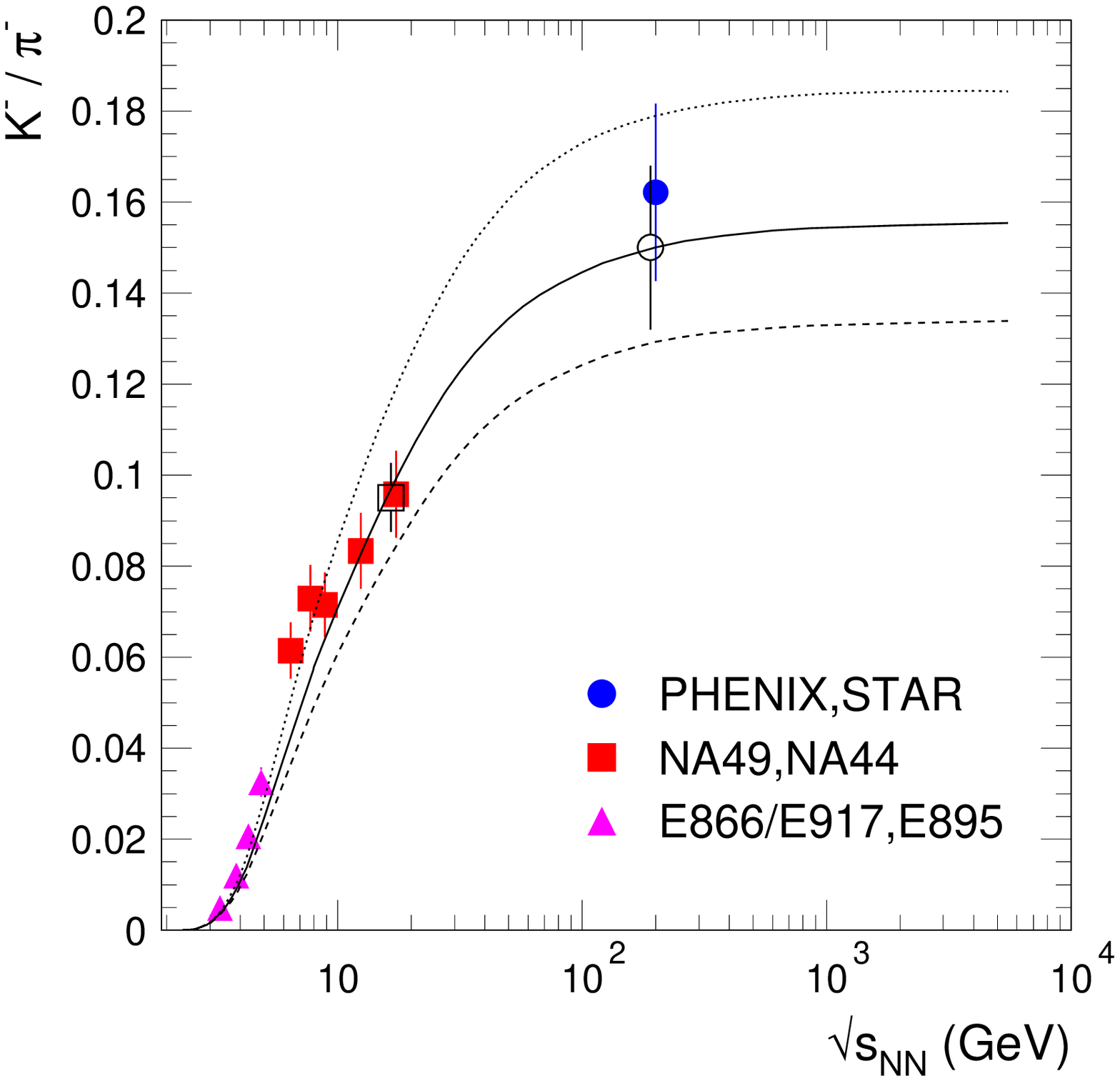}
\end{minipage}
\\
\begin{minipage}{.49\textwidth}
\vspace{-1.cm}
\centering\includegraphics[width=1.2\textwidth]{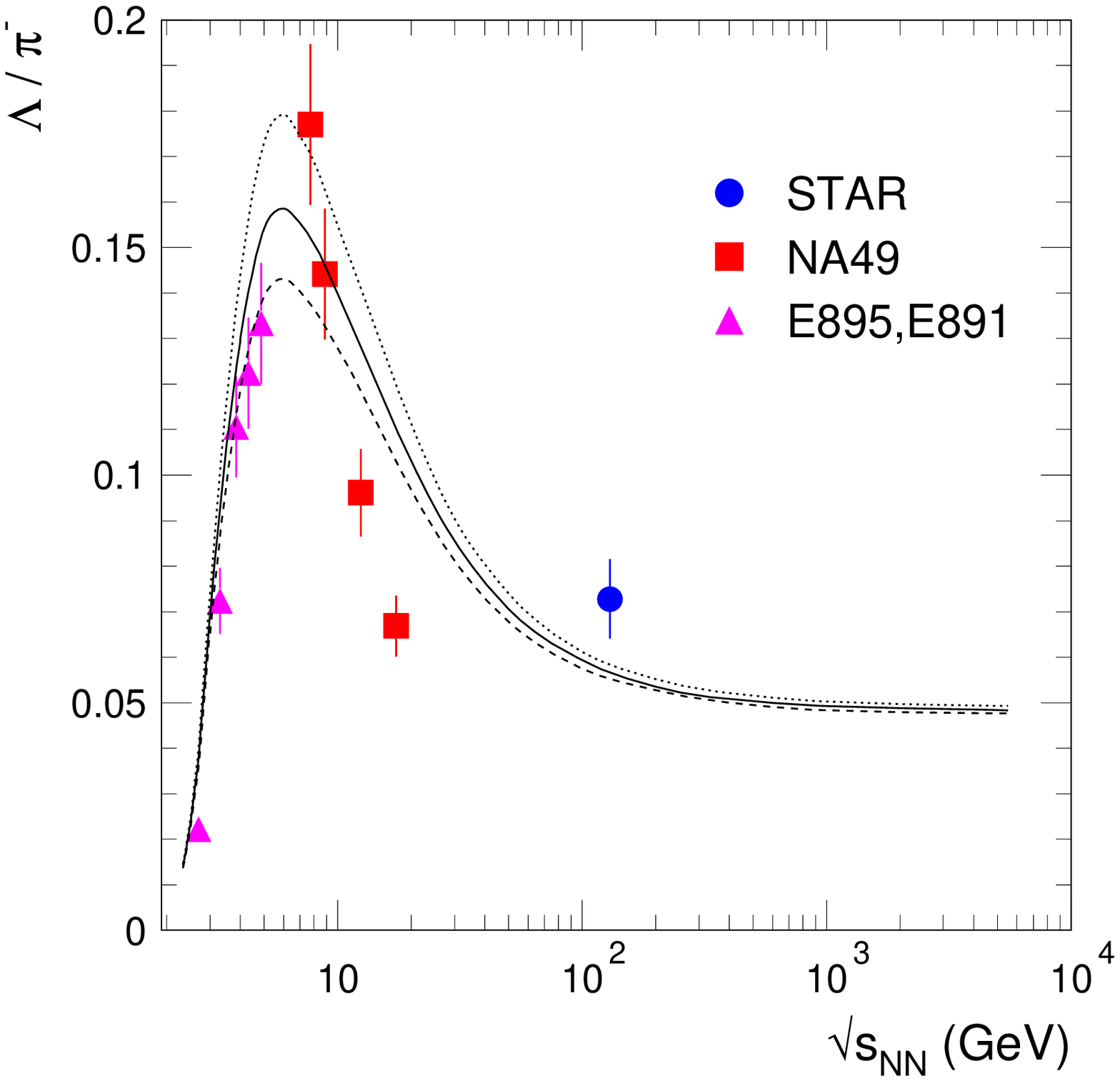}
\end{minipage} & \begin{minipage}{.49\textwidth}
\vspace{-1.cm}
\centering\includegraphics[width=1.2\textwidth]{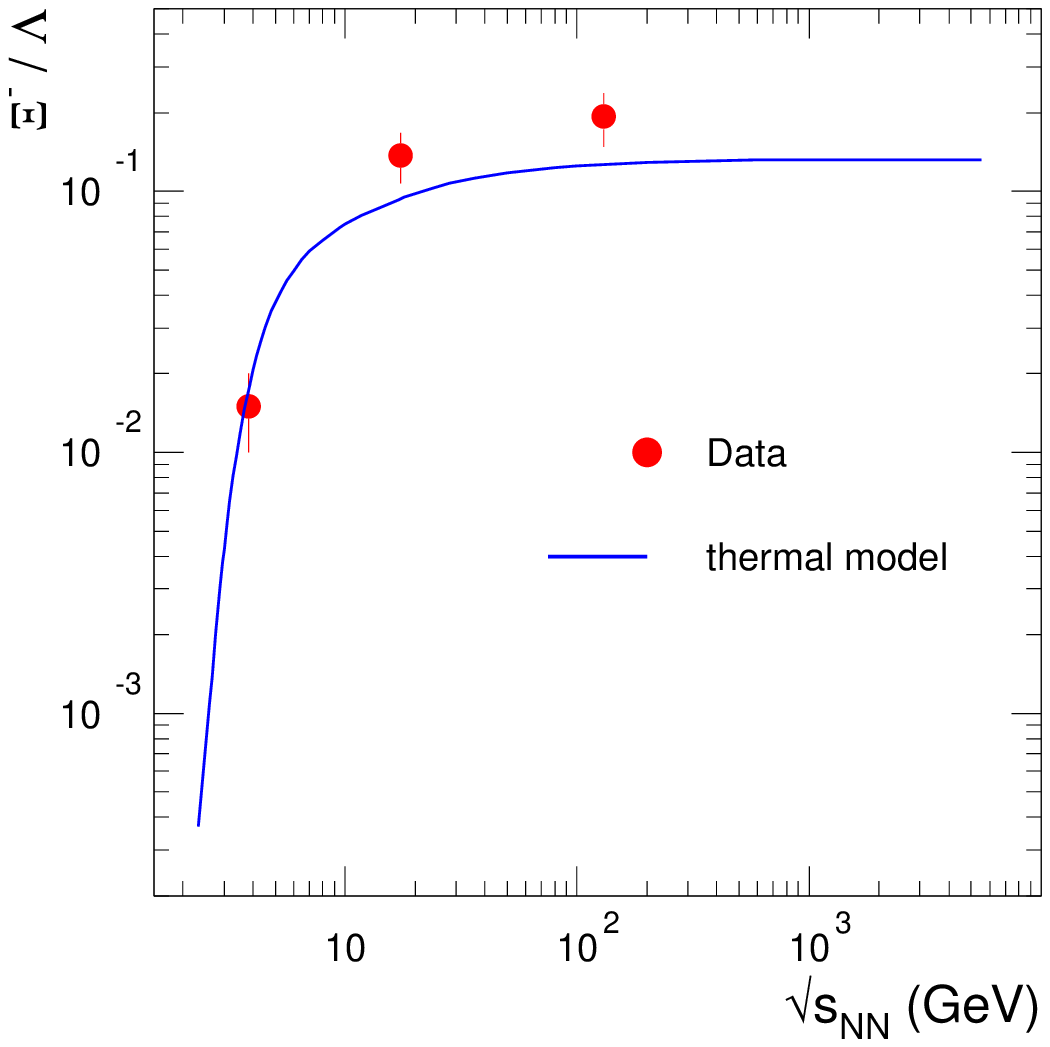}
\end{minipage} \end{tabular}
\caption{Excitation function for strange particle production, $K^\pm$ 
and $\Lambda$ yields relative to pions, $\Xi^-$ relative to $\Lambda$.
All the measured data (symbols) are for midrapidity, with the
exception of $\Lambda$ and $\Xi^-$ at AGS, for which only 4$\pi$ yields are 
available.
The lines are thermal model calculations for three cases of weak decay
reconstruction efficiencies (see text).} 
\label{aa:ks}
\end{figure}

In Fig.~\ref{aa:ks} we present excitation functions for a selection of 
strange particle yields over the whole energy range from lowest AGS to
LHC energy.
The experimental ratios $K^\pm/\pi^\pm$, $\Lambda/\pi^+$ and $\Xi^-/\Lambda$
are calculated from measurements of absolute yields of 
$\pi^\pm$ \cite{aa:e895,aa:e866a,aa:na49,aa:na44,aa:phen,aa:star},
$K^\pm$ \cite{aa:e866b,aa:na49,aa:na44,aa:phen,aa:star}, 
$\Lambda$ \cite{aa:lam} and $\Xi^-$ \cite{aa:xi}.
The errors reflect the systematic uncertainties.
These ratios are compared to thermal model calculations employing the 
parametrizations of $T$ and $\mu_b$ of Fig.~\ref{aa:tmus}.
In case of the calculations the contribution (mainly important for pion 
yields) of down-feeding from resonances (via their weak decays) is taken 
into account in three different cases, assuming that none, 50\% or all 
of the weak decays contribute to the yields. 
As one can see, the effect is significant, implying that it is very important 
that the experimental conditions (vertex cuts for selecting particles) for 
extracting the yields are well specified and taken into account in the 
model calculations. 
In a way, the extremes in weak decays reconstruction fraction shows
the range of systematic uncertainties that can arise in the comparison
of model results with experimental data, if experimental information
on feeding is ignored (or not known).

Given the accuracy of the description of multi-particle ratios presented 
above, it is not surprising that overall the model does reproduce the
experimental values rather well up to RHIC energies.
The observed discrepancies can be explained by the constant temperature
(170 MeV) used for these calculations, which is not identical (although
close) with the temperatures extracted from fits of multiparticle ratios 
shown above.
An apparent disagreement between measurements and the model calculations 
is seen concerning the energy dependence of the $K^+/\pi^+$ and 
$\Lambda/\pi^+$ ratios at SPS energies.
The origin of the rather narrow structure in the data is currently much 
debated \cite{aa:gaz}.
We note that transport models also cannot reproduce the $K^+/\pi^+$ ratio 
\cite{aa:bra}. 

The four ratios presented in Fig.~\ref{aa:ks} have a very different dependence
on energy, which reflects the evolution of the fireball at freeze-out,
dominated by the initial nucleons at low energies and by the newly created 
particles at RHIC and beyond. 
At LHC, it is expected that the fireball will consist exclusively of created 
particles.
The steep variation of the ratios at the lowest energies reflects the close
threshold for strangeness production.
The canonical suppression plays an important role as well.

\begin{figure}[htb]
\vspace{-1.cm}
\begin{tabular}{cc}
\begin{minipage}{.61\textwidth}
\centering\includegraphics[width=1.15\textwidth]{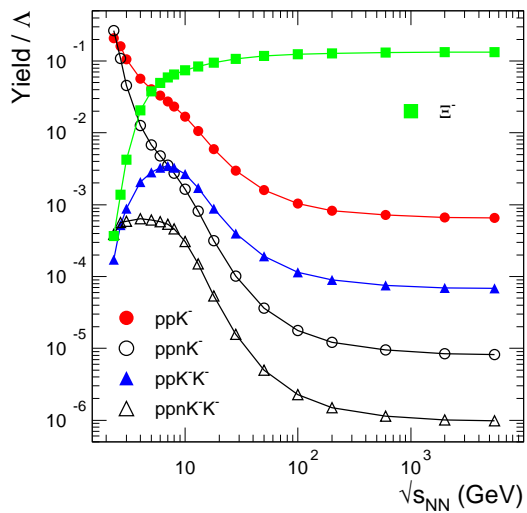}
\end{minipage} & \begin{minipage}{.35\textwidth}
\caption{Energy dependence of thermal yields of single and double $K^-$ 
clusters relative to $\Lambda$. 
The yield of $\Xi^-$ is included for reference.}
\label{aa:ksys}
\end{minipage} \end{tabular}
\end{figure}

\vspace{-.7cm}
With the $T$ and $\mu_b$ values fixed by the fits to the measured
particle ratios over a broad energy range, the thermal model has a 
good predictive power for all possible particles that can be formed 
at freeze-out.
As an example, predictions for thermal yields of $K^-$ clusters \cite{aa:yama}
relative to $\Lambda$ hyperon are shown in Fig.~\ref{aa:ksys} in comparison 
with the ratio $\Xi^-/\Lambda$.
Such exotic $K^-$ bound states have been predicted to form due to the 
strongly attractive $K^-$ potential within nuclear matter \cite{aa:yama},
but are not yet observed experimentally.
The yield of single-$K^-$ systems have large values, significantly above 
$\Xi^-$ yields, at low energies and exhibit a pronounced decrease as a 
function of energy. 
The energy dependence of double-$K^-$ systems exhibits a broad maximum around
$\sqrt{s_{NN}}\simeq$6~GeV, a region which will be covered by the future
GSI accelerator \cite{aa:gsi}.

In closing this section, we note that an open question remains concerning 
statistical model description of strongly decaying resonances (like $\rho$ 
meson and $\Delta$ baryon).
Their yields are strongly underestimated by the calculations 
\cite{aa:shu2,aa:therm}. %

\section{Charmonium and charmed hadrons} 

The importance of the so-called hard probes, among which the creation of
heavy-quarks ($c$ and $b$) have a prominent place, stems from the
fact that they are exclusively created in primary hard collisions.
Consequently, they are ideal messengers of the early stage (QGP phase) 
of the collision.
In particular the $J/\psi$ meson, which is a bound state of $c$ and $\bar{c}$
quarks, was predicted to melt in the quark-gluon plasma \cite{aa:r2}, thus 
providing a clear signature of its existence.
Although recent theoretical investigations based on lattice QCD cast doubt
on the melting at $T<1.5T_c$ \cite{aa:melt}, there is continued interest in 
quarkonia as probes of the QGP.

\begin{figure}[htb]
\vspace{-1cm}
\begin{tabular}{cc} \begin{minipage}{.62\textwidth}
\centering\includegraphics[width=1.15\textwidth]{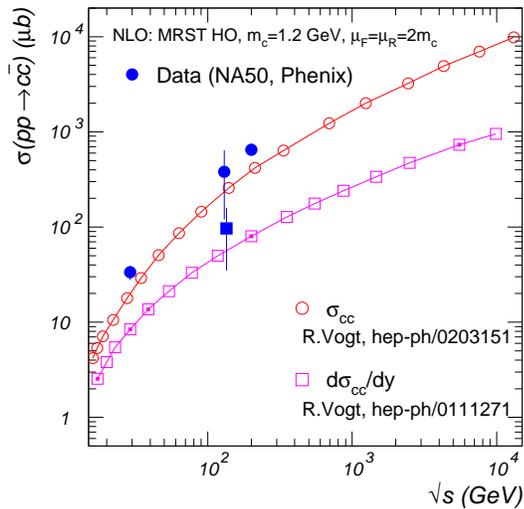}
\end{minipage} & \begin{minipage}{.35\textwidth} 
\caption{Energy dependence of the total charm cross section in elementary 
(pp) collisions. The measurements performed with nucleus-nucleus experiments 
are compared to NLO pQCD calculations \cite{aa:vogt} for the integral and 
rapidity density cross section.}
\label{aa:charm}
\end{minipage} \end{tabular}
\end{figure}

\vspace{-.5cm}
The production mechanisms of open charm (D mesons) and open beauty 
(B mesons) in elementary collisions can be well described by perturbative 
QCD (pQCD) calculations.
For instance, data on charmed meson production over a broad energy range 
was found \cite{aa:pbmy} to be in good agreement with calculations using 
the PYTHIA code (in leading order approximation, so that a scale factor of 
5 has been used in ref. \cite{aa:pbmy} to approximate the next-to-leading 
order, NLO).
Available experimental data on quarkonia production in pp collisions 
($J/\psi$ data for $\sqrt{s}$ below 100 GeV and the $\Upsilon$ family
data up to Tevatron energy, $\sqrt{s}$=1.8 TeV) have been successfully 
compared to pQCD calculations \cite{aa:gavai}.
Recent measurements of $J/\psi$ in pp collisions at RHIC are well described 
by (tuned) pQCD calculations, 
together with the measurements available at lower energies \cite{aa:sato}.

NLO pQCD calculations for total charm cross section in elementary 
collisions show clearly that the results depend significantly on the choice 
of several parameters, like the parton distribution function (PDF), 
charm quark mass (m$_c$) and renormalization and factorization constants, 
$\mu_R$ and $\mu_F$ \cite{aa:vogt}.
This dependence is the bigger the larger the energy, so it is most crucial 
for LHC energies.
A comparison of these calculations with data is presented in 
Fig.~\ref{aa:charm}. With this choice of parameters, the calculations
somewhat underpredict the measured values.
Note that all the measurements are indirect: 
at SPS the cross section was estimated from the measured Drell-Yan cross 
section in pp \cite{aa:na50b}, while at RHIC it was extracted from the 
charm contribution to the single-electron spectra measured in Au-Au 
collisions \cite{aa:phe}.

\begin{figure}[hbt]
\vspace{-1cm}
\centering\includegraphics[width=.77\textwidth]{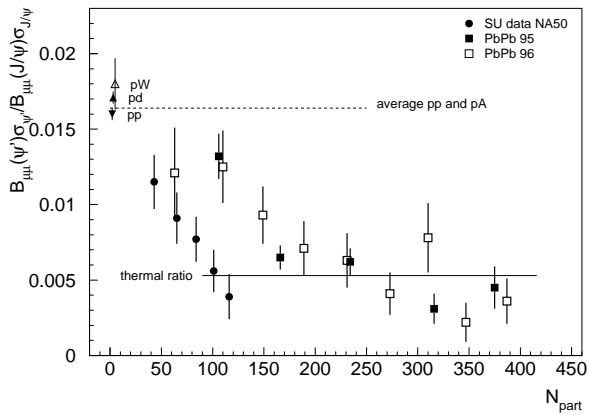}

\vspace{-1cm}
\caption{Centrality dependence of the ratio $\psi'/(J/\psi)$ 
(including branching ratios into $\mu^+\mu^-$) at SPS.}
\label{aa:psirat}
\end{figure}

In nucleus-nucleus collisions, the $J/\psi$ production at SPS is well 
measured by the NA50 collaboration \cite{aa:na50} (see also ref. 
\cite{aa:phil} for an in depth discussion).
The measured $\psi'/\psi$ ratio \cite{aa:na50pri} is independent 
of energy and is in p-A collisions the same as in pp. 
This ratio is decreasing as a function of centrality in Pb-Pb collisions, 
as seen in Fig.~\ref{aa:psirat} \cite{aa:na50pri,aa:pbm1}, and reaches,
for central collisions, a value expected for a thermal ensemble at 
$T\simeq$170~MeV \cite{aa:sor,aa:pbm1}. For an interpretation of this result,
see below.

At RHIC, the recent measurements of $J/\psi$ \cite{aa:phj} are hampered by 
very poor statistics, but high quality data are expected in the near future.
First results on open charm production at RHIC in d-Au collisions have 
just been announced \cite{aa:antai}.
In the near future, open charm cross sections will be extracted 
from recently-completed measurements in In-In collisions by NA60 
\cite{aa:na60}.
At LHC, there are good prospects to measure a complete set of charm and
bottom particles \cite{aa:hard}, in particular with the ALICE experiment.

Below we discuss the QGP fingerprints as could be unraveled through 
the model of statistical hadronization of charm quarks \cite{aa:pbm1}.
A kinetic model description of $J/\psi$ production has been independently 
developed \cite{aa:the}. 
It is equivalent from the point of view of the physical assumptions with
the model discussed here, but differs in its numerical realization.
Other approaches to statistical hadronization exist \cite{aa:gor,aa:gra},
which differ from the model discussed here, but mostly in terms of inputs,
while the outcome is qualitatively similar.
The statistical hadronization model (we follow here the outline of 
ref.~\cite{aa:aa}) assumes that all charm quarks are produced in primary 
hard collisions and equilibrate\footnote{This implies thermal, but not 
chemical equilibrium for charm quarks.} 
in the quark-gluon plasma. 
An important corollary of this assumption is that no $J/\psi$ mesons are 
preformed in the QGP, implying that the dissociation of $J/\psi$ in
QGP \cite{aa:r2} is complete. 
As noted above, recent lattice QCD calculations show that the J/$\psi$ 
mesons may not be dissociated in a deconfined medium below about $1.5T_c$ 
\cite{aa:melt}. However, it is possible that, from SPS energy on, the 
initial temperature achieved in the collision exceeds this value.

The question of charm equilibration is a difficult one, but needs to be 
addressed.
The cross sections for production of charmed hadrons are much too small 
\cite{aa:pbm4} to allow for their chemical equilibration in a hadronic gas.
But how can the apparently ``thermal'' values of the ratio $\psi'/\psi$
be reconciled with this finding ?
We assume that all charm quarks are produced in initial hard collisions,
but that open and hidden charm hadrons are formed at chemical 
freeze-out according to statistical laws. Consistent with the fact that,
at the top SPS energy and beyond, chemical freeze-out appears to be at the 
phase boundary (see previous section), the model implies that a QGP phase 
was a stage in the evolution of the fireball. 
The analysis of $J/\psi$ spectra at SPS  \cite{aa:bug} lends further support 
to the statistical hadronization picture where $J/\psi$ decouples at 
chemical freeze-out. 
A recent analysis of single-electron spectra at RHIC \cite{aa:bat} also
strenghtens the case for an early thermalization of heavy quarks.
However, in that analysis it was pointed out that both the hydrodynamical 
approach and PYTHIA reproduce the measured single-electron spectra,
although the two approaches are different in detail at low $p_t$ and 
differ manifestly at high $p_t$ ($p_t \gg$ mass of charm quark).
Another theoretical analysis \cite{aa:dok} indicates though that charm quarks 
might not thermalize quickly because of their large mass. 
All of this emphasizes the need to have high-precision direct measurements
of open charm, which could impose constraints on different interpretations. 

In  statistical models charm production needs to be treated within the 
framework of canonical thermodynamics \cite{aa:therm}. 
Thus, the charm balance equation  required during hadronization is expressed 
as:
\begin{equation}
N_{c\bar{c}}^{dir}=\frac{1}{2}g_c N_{oc}^{th}
\frac{I_1(g_cN_{oc}^{th})}{I_0(g_cN_{oc}^{th})} + g_c^2N_{c\bar c}^{th}.
\label{aa:eq1}
\end{equation}
Here $N_{c\bar{c}}^{dir}$  is the number of directly produced
$c\bar{c}$ pairs and  $I_n$ are  modified Bessel functions. In
the fireball of volume $V$  the total number of open
$N_{oc}^{th}=n_{oc}^{th}V$ and hidden  $N_{c\bar c}^{th}=n_{c\bar c}^{th}V$  
charm hadrons  are computed from their grand-canonical densities 
$n_{oc}^{th}$ and $n_{c\bar c}^{th}$, respectively. 
The densities of different particle species in the grand canonical ensemble 
are calculated following the statistical model \cite{aa:therm} introduced 
in the previous section.
All known charmed mesons and hyperons and their decays are included 
in the calculations.

The balance equation (\ref{aa:eq1}) defines a fugacity parameter $g_c$ 
that accounts for deviations of charm multiplicity  from the value that 
is expected in complete chemical equilibrium.
The yield  of open charm mesons and hyperons $i$ and of charmonia $j$ is 
obtained from:
\begin{equation}
N_i=g_cN_i^{th}\frac{{I_1(g_c N_{oc}^{th})}}{{I_0(g_c N_{oc}^{th})}}
\quad \mathrm{and} \quad N_j=g_c^2 N_j^{th}. \label{aa:eq2}
\end{equation}

The above model for charm production and hadronization can be only used 
if the number of participating nucleons $N_{part}$ is sufficiently large. 
Taking into account the measured dependence of the relative yield of 
$\psi'$ to $J/\psi$ on centrality in Pb--Pb collisions at SPS energy, 
seen in Fig.~\ref{aa:psirat}, the model appears appropriate for 
$N_{part}>$100, for which the ratio approaches the thermal value 
\cite{aa:sor,aa:pbm1}.

To calculate the yields of open and hidden charm hadrons for a
given centrality and collision energy one needs to fix a set of
parameters in Eq.(\ref{aa:eq1}) and (\ref{aa:eq2}):

i) A constant temperature of 170 MeV and a baryonic chemical potential 
$\mu_b$ according to the parametrization (\ref{eq:mu_b}) are used for 
our calculations (see Fig~\ref{aa:tmus}). 
These thermal parameters are consistent with those required to describe 
experimental data on  different hadron yields for SPS and RHIC energies.

ii) The volume of the fireball. 
We focus on rapidity density calculations
which are of relevance for the colliders, so in this case the volume 
corresponds to a slice of one rapidity unit at midrapidity, $V_{\Delta y=1}$.
It is obtained from the charged particle rapidity density $\ud N_{ch}/\ud y$, 
via the relation $\ud N_{ch}/\ud y=n_{ch}^{th}V_{\Delta y=1}$, where 
$n_{ch}^{th}$ is the charged particle density computed within the thermal 
model. 
The charged particle rapidity densities (and total yields in case of SPS, 
for which we calculate 4$\pi$ yields for a direct comparison to experimental
data) are taken from experiments at SPS and RHIC and extrapolated to 
LHC energy (as seen in Fig.~\ref{aa:nch}). 
Central collisions correspond to $N_{part}$=350.
For the centrality dependences we assume that the volume of the fireball
is proportional to $N_{part}$.

\begin{table}[htb]
\caption{Input ($\ud N_{ch}/\ud y$ and $\ud N_{c\bar{c}}^{dir}/\ud y$) 
and output ($V_{\Delta y=1}$ and $g_c$) parameters for model calculations 
at top SPS, RHIC and LHC for central collisions ($N_{part}$=350).}
\label{aa:tab1}
\begin{tabular}{c|ccc} 
$\sqrt{s_{NN}}$ (GeV) &  17.3  &  200 &  5500  \\ \hline
$\ud N_{ch}/\ud y$  &  430  &  730  &  2000  \\ 
$\ud N_{c\bar{c}}^{dir}/\ud y$  & 0.064 & 1.92  & 16.8  \\ \hline
$V_{\Delta y=1}$ (fm$^{3}$)     &  861  & 1663  & 4564 \\
$g_c$                           & 1.86  & 8.33  & 23.2 \\ 
\end{tabular}
\end{table}

iii) The yield of open charm $\ud N_{c\bar{c}}^{dir}/\ud y$ at midrapidity
(or in full volume) is taken from NLO pQCD calculations for pp collisions 
\cite{aa:vogt} and scaled to nucleus--nucleus collision via the nuclear 
overlap function, $T_{AA}$ \cite{aa:overlap}. 
For a given centrality:
\be \frac{\ud N_{c\bar{c}}^{dir}}{\ud y}(N_{part})= 
\frac{\ud \sigma({pp}\rightarrow {c\bar{c}})}{\ud y} T_{AA}(N_{part}).
\ee
The pQCD calculations with the MRST HO PDF are used here. 

The input values $\ud N_{ch}/\ud y$ and $\ud N_{c\bar{c}}^{dir}/\ud y$ 
and the corresponding volume at midrapidity and enhancement factor are 
summarized in Table~\ref{aa:tab1} for model calculations for different 
collision energies.


 We first compare predictions of the model to 4$\pi$-integrated
$J/\psi$ data at the SPS measured by NA50 collaboration \cite{aa:na50,aa:gos}. 
For the fireball total volume $V$=3070~fm$^3$ (for $N_{ch}=1533$)
the total yield of thermal open charm pairs is $N_{oc}^{th}$=0.98.
 This is to be contrasted with $N_{c\bar{c}}^{dir}$=0.137 from NLO 
calculations \cite{aa:vogt}, leading to a value of $g_c$=0.78.
Although $g_c$ is here close to unity, this obviously does not
indicate that charm production appears at chemical equilibrium, 
as the suppression factor is a strongly varying function of the
collision energy. We have already indicated that, within the time
scales available in heavy ion collisions, the chemical equilibration 
of charm  is very unlikely both in confined and deconfined media.

\begin{figure}[htb]
\begin{tabular}{cc}
\begin{minipage}{.55\textwidth}
\hspace{-1.cm}
\centering\includegraphics[width=1.1\textwidth]{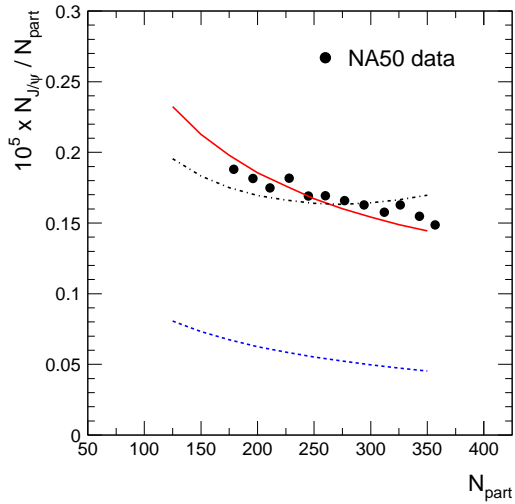}
\end{minipage} & \begin{minipage}{.4\textwidth}
\vspace{-1.cm}
\caption{The centrality dependence of $J/\psi$ production at SPS.
Model predictions are compared to 4$\pi$-integrated NA50 data 
\cite{aa:na50,aa:gos}.
Two curves for the model correspond to the values of 
$N_{c\bar{c}}^{dir}$ from NLO calculations (dashed line) and scaled up 
by a factor of 2.8 (continuous line). 
The dash-dotted curve is obtained when considering the possible NA50 
$N_{part}$-dependent charm enhancement over their extracted pp cross 
section \cite{aa:na50b} (see text).}
\label{aa:fig1}
\end{minipage} 
\end{tabular}
\end{figure}

In Fig.~\ref{aa:fig1} we show the comparison between the results
of our model and NA50 data for two different values of
$N_{c\bar{c}}^{dir}$: from NLO calculations \cite{aa:vogt} and
scaled up by a factor of 2.8. 
Using the NLO cross sections for charm production scaled by the 
nuclear overlap function, the model understimates the measured yield.
To explain the overall magnitude of the data, we need to 
increase the $N_{c\bar{c}}^{dir}$ yield  by a factor of 2.8 as compared 
to NLO calculations. We mention in this context that the observed
\cite{aa:na50b} enhancement of the di-muon yield at intermediate
masses has been interpreted as a possible indication for an anomalous 
increase of the charm production cross section. 
A third calculation (resulting in the dash-dotted line in 
Fig.~\ref{aa:fig1}) is using the NLO cross section scaled-up by 1.6,
which is the ratio of the open charm cross section estimated by NA50 for 
pp collisions at 450 GeV/c \cite{aa:na50b} and the present NLO values.
For this case the $N_{part}$ scaling is not the overlap function, but
is taken according to the measured di-muon enhancement as a function
of $N_{part}$  \cite{aa:na50b}.
The resulting $J/\psi$ yields from the statistical model are on
average in agreement with the data, albeit with a flatter centrality 
dependence than by using the nuclear overlap function. 
Thus our charm enhancement factor of 2.8 needed to explain the $J/\psi$ 
data is very similar to the factor needed to explain the intermediate mass 
dilepton enhancement assuming that it arises exclusively from charm
enhancement \cite{aa:na50b}.
 We note, however, that other plausible explanations exist of the 
observed enhancement in terms of thermal radiation \cite{aa:rapp}.

\begin{table}[htb]
\caption{Mid-rapidity densities for open and hidden charm hadrons,
calculated for central collisions ($N_{part}$=350) at SPS, RHIC and LHC.}
\label{aa:tab2}
\begin{tabular}{c|ccc} 
$\sqrt{s_{NN}}$ (GeV) &  17.3  &  200 &  5500  \\ 
\hline
D$^+$   & 0.010 & 0.404 & 3.56 \\ 
D$^-$   & 0.016 & 0.420 & 3.53 \\ 
D$^0$   & 0.022 & 0.888 & 7.80 \\ 
$\bar \mathrm{D}^0$ & 0.035 & 0.928 & 7.82 \\ 
D$^{*+}$ & 0.009 & 0.374 & 3.30 \\ 
D$^{*-}$ & 0.015 & 0.393 & 3.30 \\ 
D$_s^+$  & 0.012 & 0.349 & 2.96 \\ 
D$_s^-$  & 0.009 & 0.338 & 2.95 \\ 
${\Lambda_c}$      & 0.014 & 0.153 & 1.16 \\ 
${\bar \Lambda_c}$ & 0.0012 & 0.117 & 1.15 \\ 
${J/\psi}$  &  2.55$\cdot$10$^{-4}$ & 0.011 & 0.226 \\ 
${\psi'}$ & 0.95$\cdot$10$^{-5}$ & 3.97$\cdot$10$^{-4}$ & 8.46$\cdot$10$^{-3}$ 
\\ 
\end{tabular}
\end{table}

We turn now to discuss our model predictions for charmonia and
open charm production at collider energies and compare them with
the  results obtained at SPS.  
Notice that from now on we focus on rapidity densities,
which are the relevant observables at the colliders.
In Table~\ref{aa:tab2} we summarize the yields for a selection of hadrons 
with open and hidden charm. All predicted yields increase strongly with 
beam energy, reflecting the increasing charm cross section and the
concomitant importance of statistical recombination. 
Also, ratios of open charm hadrons evolve with increasing energy, 
reflecting the corresponding decrease in the charm chemical potential.
Very recent measurements of open charm in d-Au collisions at RHIC 
\cite{aa:antai} yield the ratio (D$^{*+}$+D$^{*-}$)/(D$^0$+$\bar \mathrm{D}^0$)
of 0.40$\pm$0.09, which is in a good agreement to the model prediction 
of 0.42.

\begin{figure}[hbt]
\begin{tabular}{cc}\begin{minipage}{.55\textwidth}
\centering\includegraphics[width=1.\textwidth]{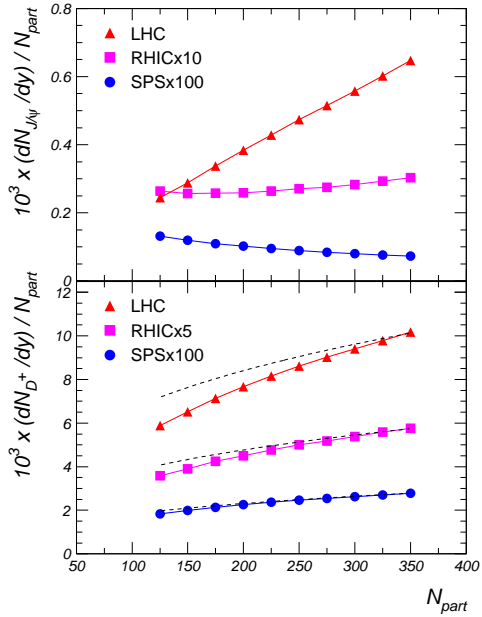}
\end{minipage} & \begin{minipage}{.4\textwidth}
\caption{Centrality dependence of rapidity densities of $J/\psi$ 
(upper panel) and D$^+$ (lower panel) mesons per $N_{part}$ at SPS,
RHIC and LHC.  Note the scale factors for RHIC and SPS energies.  The
dashed lines in the lower panel represent $N_{part}^{1/3}$ dependences
normalized for $N_{part}$=350.  }
\label{aa:fig2}
\end{minipage} \end{tabular}
\end{figure}

Model predictions for the centrality dependence of $J/\psi$ and D$^+$
rapidity densities normalized to $N_{part}$ are shown in
Fig.~\ref{aa:fig2}.  The results for $J/\psi$ mesons exhibit, in
addition to the dramatic change in magnitude, a striking change in the
shape of the centrality dependence. In terms of the model this change
is a consequence of the transition from the canonical to the
grand-canonical regime.  For  D$^+$-mesons, the expected
approximate scaling of the ratio D$^+$/$N_{part}$ 
$\propto N_{part}^{1/3}$ (dashed lines in Fig.~\ref{aa:fig2}) is only roughly
fulfilled due to departures of the nuclear overlap function from the
simple $N_{part}^{4/3}$ dependence.

The results summarized in Table~\ref{aa:tab2} and shown in
Fig.~\ref{aa:fig2} obviously depend on two input parameters,
$\ud N_{ch}/\ud y$ and $\ud N_{c\bar{c}}^{dir}/\ud y$. 
For LHC energy, neither one of these parameters is well known. 
An increase of charged particle  multiplicities by up to a factor 
of three beyond  our ``nominal'' value $\ud N_{ch}/\ud y$=2000 for
central collisions is conceivable.
However, due to quite large uncertainties on the  amount of shadowing 
at LHC energy, these results may be still modified. The yield of 
$\ud N_{c\bar{c}}^{dir}/\ud y$ is also not well known at LHC energy. 
Although these uncertainties affect considerably the magnitude of the
predicted yields,  their centrality dependence remains
qualitatively unchanged: the yields per participant are increasing
functions of $N_{part}$.  We also note here that, while detailed
predictions differ significantly, qualitatively similar results
(see ref.~\cite{aa:hard}) have been obtained for a kinetic model study 
of $J/\psi$ production at the LHC.

\begin{figure}[hbt]
\begin{tabular}{cc}\begin{minipage}{.55\textwidth}
\centering\includegraphics[width=.97\textwidth]{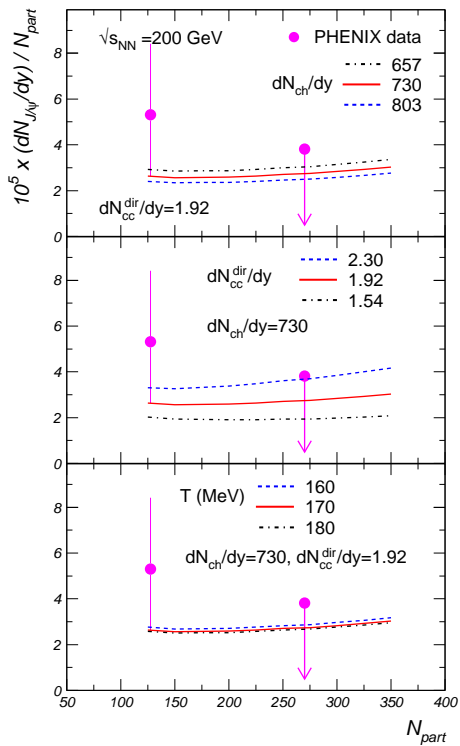}
\end{minipage} & \begin{minipage}{.4\textwidth}
\caption{Centrality dependence of rapidity densities of $J/\psi$
mesons at RHIC. Upper panel: sensitivity to $\ud N_{ch}/\ud y$; 
middle panel: sensitivity to $\ud N_{c\bar{c}}^{dir}/\ud y$; 
lower panel: sensitivity to $T$.  
The calculations are represented by lines.
The dots are experimental data from the PHENIX collaboration \cite{aa:phj}.
Note that the point for the central collisions is the upper limit
extracted by PHENIX for 90\% C.L. \cite{aa:phj}.}
\label{aa:fig3}
\end{minipage} \end{tabular}
\end{figure}

\vspace{-.5cm}
In Fig.~\ref{aa:fig3} we present the predicted centrality
dependence of the $J/\psi$ rapidity density normalized to
$N_{part}$ for RHIC energy ($\sqrt{s_{NN}}$=200 GeV).  
The three panels show its sensitivity on $\ud N_{ch}/\ud y$, 
$\ud N_{c\bar{c}}^{dir}/\ud y$, and (freeze-out) temperature $T$.  
The calculations are compared to experimental results of
the PHENIX Collaboration \cite{aa:phj}.
 The experimental data have been rescaled according to our procedure 
to calculate $N_{part}$ and the number of binary collisions, $N_{coll}$.  
Within the still  large experimental error bars, the measurements
agree with our model predictions.  In Fig.~\ref{aa:fig3} only the
statistical errors of the mid-central data point are plotted. 
The systematic errors are also large \cite{aa:phj}.
A stringent test of the present model can
only be made when high statistics $J/\psi$ data are available. 

We turn now to a more detailed discussion of the sensitivity of
our calculations to the various input parameters as quantified  in
Fig.~\ref{aa:fig3}. First we consider  the influence of a $10\%$
variation of $\ud N_{ch}/\ud y$ on the centrality dependence of
$J/\psi$ yield.  Note that the total experimental uncertainty of
$\ud N_{ch}/\ud\eta$ (which is for the moment the measured
observable for most experiments) at RHIC is below 10\%.
The sensitivity on the $\ud N_{ch}/\ud y$ values stems from the volume
into which the (fixed) initial number of charm quarks is
distributed. The smaller the particle multiplicities and thus also
the fireball volume, the more probable it is for charm quarks and
antiquarks to combine and form quarkonia. That is why one sees, in
the top panel in Fig.~\ref{aa:fig3}, that the $J/\psi$ yield is
increasing with decreasing charge particle multiplicity.

The sensitivity of the predicted $J/\psi$ yields on 
$\ud N_{c\bar{c}}^{dir}/\ud y$ is  also straightforward. The larger this 
number is in a fixed volume the larger is the yield of charmed hadrons. 
In case of charmonia the dependence on $\ud N_{c\bar{c}}^{dir}/\ud y$ 
is non-linear due to their double charm quark content, as reflected 
by the factor $g_c^2$ in equation (\ref{aa:eq2}). 
To illustrate the sensitivity of the model predictions on 
$\ud N_{c\bar{c}}^{dir}/\ud y$, we exhibit the results of a 20\% variation
with  respect to the value  given in Table~\ref{aa:tab1}. The open
charm cross section is not yet measured at RHIC. However, some
indirect measurements can be well reproduced, within the
experimental errors, by PYTHIA calculations using a p--p charm
cross sections scaled with the number of collisions $N_{coll}$ of
650 $\mu$b \cite{aa:phe}. The corresponding value at
$\sqrt{s_{NN}}$=130 GeV is 330 $\mu$b \cite{aa:phe}. For comparison,
the NLO pQCD values we are using are 390 and 235 $\mu$b, 
respectively. Despite the still large experimental uncertainties, 
this discrepancy needs to be understood.  
We note that, dependent on the input parameters used in the NLO calculations
\cite{aa:vogt}, possible variations of the open charm production 
cross section for the RHIC energy are of the order of $\pm$20\%. 
In terms of our model this variation corresponds 
to about a $\pm$30\% change of the $J/\psi$ yield, which is also
centrality dependent (see middle panel in Fig.~\ref{aa:fig3}). 
If we use the PHENIX p--p cross section of 650 $\mu$b, the
calculated yield is a factor 2.5 larger for $N_{part}$=350 and
increases somewhat stronger with centrality. 
As apparent in Fig.~\ref{aa:fig3}, the predictive power of this model, 
or of any similar model, relies heavily on the accurate knowledge of 
the charm production cross section.  
A simultaneous description of the centrality dependence of open charm
together with $J/\psi$ production is, in this respect, mandatory to test 
the  concept of the statistical origin of open and hidden charm 
hadrons in heavy ion collisions at relativistic energies.

The apparent weak dependence of $J/\psi$ yield on  freeze-out
temperature, seen in  Fig.~\ref{aa:fig3}, may be surprising. 
In our model this result is a consequence of  the charm balance equation
(\ref{aa:eq2}).  The temperature variation leads, obviously, to a
different number of thermally produced charmed hadrons, but this is 
compensated by the $g_c$ factor.
The approximate temperature dependence of $g_c$ and the $J/\psi$ yield are:
\begin{equation}
g_c(T) \sim 1/N_D^{th} \sim e^{\frac{m_D}{T}}, \quad
N_{J/\psi}(T)=g_c^2 N_{J/\psi}^{th} \sim e^{\frac{2m_D-m_{J/\psi}}{T}}
\end{equation}
As a result of the small mass difference in the exponent  the
$J/\psi$ yield exhibits  only a weak sensitivity on  $T$. This
is in contrast to the purely thermal case where the yield  scales
with $\exp(-m_{J/\psi}/T)$.
The only exception is the ratio $\psi'/J/\psi$, which is obviously 
identical in the statistical hadronization scenario and in the thermal 
model and coincides, for T$\simeq$170 MeV, with the measured value at 
SPS (see Fig.~\ref{aa:psirat}).

\begin{figure}[hbt]
\begin{tabular}{cc}\begin{minipage}{.55\textwidth}
\hspace{-1cm}
\centering\includegraphics[width=1.1\textwidth]{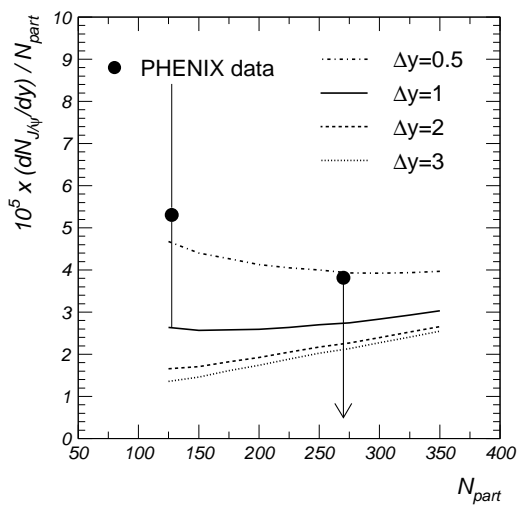}
\end{minipage} & \begin{minipage}{.4\textwidth}
\vspace{-1cm}
\caption{Centrality dependence of rapidity densities of $J/\psi$
mesons at RHIC for different rapidity window sizes. 
The lines are calculations, the dots are experimental data from 
PHENIX collaboration \cite{aa:phj} (the point for the central collisions 
is the upper limit for 90\% C.L.).
} \label{aa:fig4}
\end{minipage} \end{tabular}
\end{figure}

Most of our results presented above are obtained considering a one
unit rapidity window at midrapidity, while results for the full volume
were presented only for the SPS. Unlike the kinetic model of
Thews et al. \cite{aa:the}, our model does not contain
dynamical aspects of the coalescence process.  However, in our
approach, the width of the rapidity window does influence the
results in the canonical regime. For the grand-canonical case,
attained only at LHC energy, there is no dependence on the width
of the rapidity window, due to a simple cancellation between the
variation of the volume, proportional to the rapidity slice in
case of a flat rapidity distribution, and the variation of
$N_{c\bar{c}}^{dir}$, also proportional to the width of the
rapidity slice.  In Fig.~\ref{aa:fig4} we present the centrality
dependence of $J/\psi$  rapidity densities for RHIC energy  and
for different rapidity windows $\Delta$y from 0.5 to 3. The
dependence on $\Delta$y resembles that of the kinetic model
\cite{aa:the}, but is less pronounced. The available  data  are not
yet precise enough to rule out any of the scenarios considered.
However, for the kinetic model, the cases of small $\Delta$y 
seem to be ruled out by the present PHENIX data.
We stress in this context that the size of the $\Delta$y window has 
a potentially large impact on the  results at  SPS energy. It is
conceivable that no charm enhancement is needed to explain the
data if one considers a sufficiently narrow rapidity window
for the statistical hadronization.

\vspace{.5cm}
\begin{table}[htb]
\begin{tabular}{cc}\begin{minipage}{.55\textwidth}
\begin{tabular}{lcc}
\hline
Particle ~~        & {Statistical}   &  ~~~{pQCD NLO}~\\ 
 ~~                & {hadronization} &  ~~  \\ 
\hline
~~~$\mathrm{D}^+$       & {0.228} & {0.155} \\
~~~$\mathrm{D}^-$       & {0.226} & {0.146} \\ 
~~~$\mathrm{D}^+_s$     & {0.190} & {0.095} \\ 
~~~$\mathrm{D}^-_s$     & {0.189} & {0.089} \\ 
~~~$\Lambda_c$          & {0.074} & {0.086} \\ 
~~~$\bar{\Lambda}_c$    & {0.074} & {0.062} \\ 
\hline
\end{tabular}
\end{minipage} & \begin{minipage}{.42\textwidth}
\caption{Ratios of midrapidity densities for open charm hadrons relative to
($\mathrm{D}^0+\mathrm{\bar{D}}^0$), calculated for central collisions at LHC. 
The results of the statistical hadronization model are compared to NLO pQCD 
calculations \cite{aa:hard}.}
\label{aa:tab3}
\end{minipage} \end{tabular}
\end{table}

 \begin{figure}
\vspace{-.5cm}
\centering\includegraphics[width=.63\textwidth]{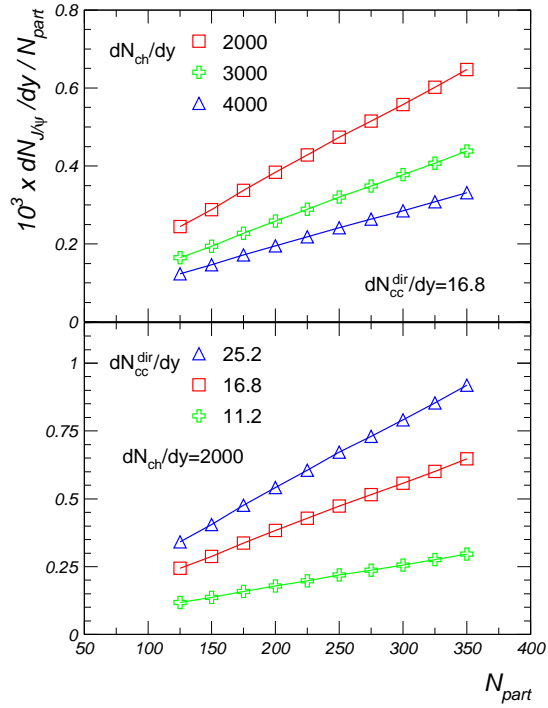} 
\caption{Centrality dependence of rapidity densities of $J/\psi$ 
(per $N_{part}$) at LHC. Upper panel: sensitivity to $\ud N_{ch}/\ud y$, 
lower panel: sensitivity to $\ud N_{c\bar{c}}^{dir}/\ud y$. }
\label{aa:figx}
\end{figure}

Ratios of mid-rapidity densities for open charm hadrons relative to
($\mathrm{D}^0+\mathrm{\bar{D}}^0$) are presented in Table~\ref{aa:tab3}
for central collisions at LHC.
The statistical hadronization model results are compared to NLO pQCD 
calculations \cite{aa:hard}.
In case of pQCD, the production of charm is identical to the elementary
case, namely charm quark production in hard processes. 
Sizeable differences (up to a factor of 2, in case of D$_s$ mesons) 
are seen. The measurements will certainly be able to distinguish between
the two scenarios.

In Fig.~\ref{aa:figx} we present the statistical hadronization model
results on rapidity densities of $J/\psi$ per $N_{part}$ for the LHC energy.
We study the sensitivity on the two input parameters that are not well 
known at LHC, $\ud N_{ch}/\ud y$ and $\ud N_{c\bar{c}}^{dir}/\ud y$.
Within the variations considered here (up to a factor 2 larger particle 
multiplicities and a $\pm$50\% in the charm cross section) the changes in 
the yields are considerable, but the dependences on $N_{part}$ remain 
the same, making this a rather solid prediction for LHC.

\begin{figure}[hbt]
\begin{tabular}{cc}\begin{minipage}{.58\textwidth}
\hspace{-.5cm}
\centering\includegraphics[width=1.05\textwidth,height=1.1\textwidth]{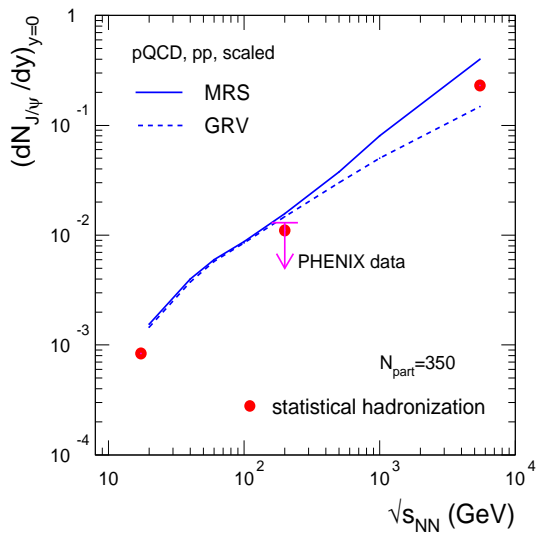}
\end{minipage} & \begin{minipage}{.35\textwidth}
\vspace{-.8cm}
\caption{Excitation function of $J/\psi$ production in central
nucleus-nucleus collisions.
The symbols are statistical hadronization model calculations.
The lines are pQCD calculations for pp collisions \cite{aa:gavai}, 
scaled for $N_{part}$=350, for two PDFs.
The arrow denotes the experimental value \cite{aa:phj}.
} \label{aa:figj}
\end{minipage} \end{tabular}
\end{figure}

The excitation function of $J/\psi$ production (rapidity densities) 
is shown in Fig.~\ref{aa:figj}. 
The statistical hadronization model results are compared to pQCD calculations 
for pp collisions \cite{aa:gavai}, scaled for $N_{part}$=350, for two PDFs.
Note that the PDFs, as well as the other inputs of the pQCD calculations 
\cite{aa:gavai} are different from those used to extract the charm cross 
section \cite{aa:vogt} which is an input to the statistical hadronization model.
In any case, the yields are comparable in the two cases, implying that fine 
tuning of the input values will be needed to be able to distinguish between
the two scenarios. As in the case of the total charm cross section 
\cite{aa:vogt}, the dependence of $J/\psi$ production on the PDF choice 
is evident for the higher energies (LHC).

The  results presented  above were obtained under the assumption
of statistical hadronization of quarks and gluons.
We have assumed that charm quarks are entirely produced via 
primary hard scattering  and thermalized in the QGP. 
No secondary production of charm in the initial and final state was 
included in our calculations.
Final state effects like nuclear absorption of $J/\psi$ \cite{aa:r2} 
are also neglected.
First RHIC data on $J/\psi$ production support the current predictions, 
although the experimental errors are for the moment too large to allow 
firm conclusions. 
Also the RHIC results on open charm lend a strong support for this model.
The statistical coalescence implies travel of charm quarks over significant 
distances e.g. in a QGP. If the model predictions will describe consistently 
precision data this would be a clear signal for the presence of a deconfined 
phase.
We emphasize that the predictive power of this (and any similar) model relies
heavily on the accuracy of the charm cross section, which is yet to be 
directly measured in nucleus-nucleus collisions.  

\section{Outlook}

The field of ultrarelativistic nucleus-nucleus collisions has reached the
stage of precision measurements, which are able to provide fundamental
information on the strongly interacting matter at high temperature
and (energy) density and in particular on the quark-gluon plasma.
It is now clear that such complex knowledge can only be achieved by a set 
of multi-faceted and complementary studies and that no one single observable 
is sufficient to characterize fully the properties  of the QGP phase. 
From what we have briefly reviewed here it is clear that the global
characterization of the collision has been convincingly achieved and, 
subject to further refinements, establishes beyond doubt that the conditions
for the creation of QGP have been attained.
We have shown that the study of particle ratios provide unique insight 
on the QCD phase diagram, while the study of charm hadrons
provides a valuable glimpse into the QGP.

We look forward to new high statistics and precision experiments from the SPS
and, in particular,  RHIC. From 2007 on the high temperature region of the
phase diagram will be investigated in detail at the LHC with the dedicated
ALICE experiment as well as within the ATLAS and CMS collaborations. The next
serious attack on the high density-moderate temperature regime will be
addressed further in the future with the new GSI accelerator
facility. Interesting times are ahead!

\end{document}